\documentclass[10pt]{article}
\usepackage{amsmath,amsfonts,amssymb}
\usepackage{hyperref}
\usepackage{graphicx}
\usepackage{subfigure}
\usepackage{multicol}
\def\beq{\begin{equation}}
\def\eeq{\end{equation}}
\def\barr{\begin{array}}
\def\earr{\end{array}}
\def\dis{\displaystyle}

\def\tev{\, {\rm TeV}}
\def\gev{\, {\rm GeV}}
\def\mpl{\overline{M}_{Pl}}
\usepackage{color}
\usepackage{ulem}

\definecolor{darkgreen}{cmyk}{1,0,1,0.4}

\def\soutps{\bgroup\markoverwith{\textcolor{darkgreen}{\rule[0.5ex]{2pt}{0.4pt}}}\ULon}

\def\lapp{\mathrel{\rlap{\raise.5ex\hbox{$<$}}
                    {\lower.5ex\hbox{$\sim$}}}}
\def\gapp{\mathrel{\rlap{\raise.5ex\hbox{$>$}}
                    {\lower.5ex\hbox{$\sim$}}}}

\begin{document}

\begin{center} 
{\large \bf Gravitons in multiply warped scenarios - at 750 GeV and beyond
} \\
\vspace*{1cm} 
\renewcommand{\thefootnote}{\fnsymbol{footnote}} 
{ {Mathew Thomas Arun${}^1$} and {Pratishruti Saha${}^{2,3}$} 
} \\ 
\vspace{10pt} 
${}^{1}$ {\it Department of Physics and Astrophysics, University of
Delhi, Delhi 110 007, India}
 
${}^{2}$ {\it Physique des Particules, 
            Universit\'e de Montr\'eal,
            C.P. 6128, succ. centre-ville, \\ Montr\'eal, QC,
            Canada H3C 3J7.}

${}^{3}$ {\it Harish-Chandra Research Institute, 
            Chhatnag Road, Jhunsi,
            Allahabad - 211019, India.}
\normalsize 
\end{center} 

%%%%%%%%%%%%%%%%%%%%%%%%%%%%%%%%%%%%%%%%%%%%%%%%%%%%%%%%%%%%%%%%%%%%%%%
% Abstract
%%%%%%%%%%%%%%%%%%%%%%%%%%%%%%%%%%%%%%%%%%%%%%%%%%%%%%%%%%%%%%%%%%%%%%%
\begin{abstract}
The search for extra dimensions has so far yielded no positive results
at the LHC. Along with the discovery of a 125 $\gev$ Higgs boson, this
implies a moderate degree of fine tuning in the parameter space of the
Randall-Sundrum model. Within a 6-dimensional warped compactification
scenario, with its own interesting phenomenological consequences, the
parameters associated with the additional spatial direction can be
used to eliminate the need for fine tuning. We examine the constraints
on this model due to the 8 $\tev$ LHC data and survey the parameter
space that could be probed at the 14 $\tev$ run of the LHC. We also
identify the region of parameter space that is consistent with the
recently reported excess in the diphoton channel in the 13 $\tev$
data.  Finally, as an alternative explanation for the observed excess,
we discuss a scenario with brane-localized Einstein-Hilbert terms with
Standard Model fields in the bulk.
\end{abstract}
%%%%%%%%%%%%%%%%%%%%%%%%%%%%%%%%%%%%%%%%%%%%%%%%%%%%%%%%%%%%%%%%%%%%%%%

%%%%%%%%%%%%%%%%%%%%%%%%%%%%%%%%%%%%%%%%%%%%%%%%%%%%%%%%%%%%%%%%%%%%%%%
% Introduction
%%%%%%%%%%%%%%%%%%%%%%%%%%%%%%%%%%%%%%%%%%%%%%%%%%%%%%%%%%%%%%%%%%%%%%%
\section{Introduction}
\label{sec:intro}

The warped geometry model proposed by Randall and Sundrum (RS)~\cite{RS} 
is one of the many models in the literature that offer a 
resolution of the well-known naturalness problem. This model
is particularly promising because 
$(i)$ it resolves the gauge hierarchy problem 
with large extra dimensions); $(ii)$ the modulus of the extra dimension
can be stabilized to a desired value by well-understood mechanisms, 
e.g. the one due to Goldberger and Wise ~\cite{GW1}, and $(iii)$ a similar 
warped solution can be obtained from a more fundamental theory like string 
theory where extra dimensions appear naturally~\cite{Green}. 
As a result, several search strategies at the LHC were designed 
specifically~\cite{diboson,EXO-12-045,dilepton,ATLASDec2015,CMSDec2015} to 
detect signatures of these warped extra dimensions through the decays of 
Kaluza-Klein (KK) excitations of the graviton which appear at the $\tev$ scale in this model. 
The results, so far, have been negative, with the ATLAS 
Collaboration~\cite{ATLAS2015} setting a lower bounds of 2.66 (1.41) $\tev$ on the mass of the 
lightest KK excitation of the graviton for a coupling of $k_5/\mpl$ = 0.1 (0.01). 
The limits obtained by the CMS Collaboration are similar~\cite{EXO-12-045}.
This result, together with the very restrictive nature of the RS model, 
necessitates a fine-tuning of 2-3 orders of magnitude in order to 
explain the observed Higgs mass of 125 $\gev$. The model will 
become even more fine-tuned if the bounds are pushed higher 
during Run 2 of the LHC.

Some of these difficulties can be eliminated by considering more generalized 
versions of the RS model. Several such models already exist in the 
literature~\cite{RSextn_previous,6D-Salvio}. In this work, we focus on one such model 
which features two extra spatial dimensions with the warping in the two 
directions being intertwined~\cite{dcssg}. The graviton spectrum in this model
was worked out in Ref.~\cite{Arun:2014dga}. While the experiments at the LHC 
focus entirely on 5-dimensional models, we re-interpret these 
bounds ~\cite{ATLAS2015} and predictions for Run 2 ~\cite{Das:2014tva}, 
in the context of this multiply warped brane world model.

Interestingly, both the CMS and the ATLAS Collaborations have 
reported~\cite{CMSDec2015,ATLASDec2015} a small excess of events in the 
$p p \to \gamma \gamma$ channel in the region near 
$m_{\gamma\gamma}$ = 750 $\gev$. While this excess can be explained in a 
whole host of scenarios~\cite{diphoton_deluge}, the CMS collaboration has 
also attempted a RS-graviton interpretation\footnote{In a later update, ATLAS also reported a spin-2 analysis}. 
However, even with a small value of the effective coupling, namely $k_5/\mpl$ = 0.01,  
the ensuing cross-section is much larger than the observed excess.
Smaller values of the coupling only occur in an unviable region of parameter space where one must either allow    
a large hierarchy between the moduli of the extra dimension and the scale of 
gravity, or, admit greater degrees of fine-tuning in order to resolve the 
gauge-hierarchy problem. As we shall show, this is not the case with the 
6-dimensional scenario where smaller values of the effective coupling
arise quite naturally. Based on these considerations, the 6-dimensional model
would appear to be a more appealing explanation of the observed signal.
However, the statistical significance of the observation is, as yet, too small 
for any strong claims to be made.

The remainder of this paper is organized as follows : 
we briefly describe the model in Section 2 and 
its graviton spectrum in Section 3; 
the limits on the parameter space are examined in Section 4, 
the compatibility of the model parameter space and an alternative model 
that could be in corroboration with the 
observed diphoton excess are discussed in Section 5; and finally, we conclude in Section 6.

%%%%%%%%%%%%%%%%%%%%%%%%%%%%%%%%%%%%%%%%%%%%%%%%%%%%%%%%%%%%%%%%%%%%%%%
% End of Introduction
%%%%%%%%%%%%%%%%%%%%%%%%%%%%%%%%%%%%%%%%%%%%%%%%%%%%%%%%%%%%%%%%%%%%%%%

%%%%%%%%%%%%%%%%%%%%%%%%%%%%%%%%%%%%%%%%%%%%%%%%%%%%%%%%%%%%%%%%%%%%%%%
% The Model
%%%%%%%%%%%%%%%%%%%%%%%%%%%%%%%%%%%%%%%%%%%%%%%%%%%%%%%%%%%%%%%%%%%%%%%
\section{6 Dimensions, 4 Branes and Nested Warping}
\label{sec:model}

The metric for the 6-dimensional 
space-time with successive orbifoldings viz. 
$M^{1,5}\rightarrow[M^{1,3}\times S^1/Z_2]\times S^1/Z_2$,
is defined as~\cite{dcssg}
\begin{equation}
ds^2_6= b^2(z)[a^2(y)\eta_{\mu\nu}dx^{\mu}dx^{\nu}+R_y^2dy^2]+r_z^2dz^2 \ .
\label{eq:metric}
\end{equation}  
Here, $y, z \in [0,\pi]$ are angular coordinates representing
the compactified directions, and $R_y$ and $r_z$ the respective moduli. 
Each of the four orbifold fixed points 
($y = 0$, $z = \pi$, $z = 0$ and $z = \pi$) are associated with 
4-branes endowed with a 
localized energy density $V_i$. 
The total gravity action is, thus, 
\begin{equation}
\barr{rcl}
{\cal{S}} = \dis \int d^4x \, dy \, dz \sqrt{-g_6} \, (M_{6}^4R_6-\Lambda) 
  &+ \dis \int d^4x \, dy \, dz \sqrt{-g_5}\, 
      [V_1(z) \, \delta(y)+V_2(z) \, \delta(y-\pi)] \\
&+ \dis \int d^4x \, dy \, dz \sqrt{-\tilde g_5} \, 
     [V_3(y) \, \delta(z)+V_4(y) \, \delta(z-\pi)] \ ,
\earr
\label{eq:action}
\end{equation}
where $M_6$ and $\Lambda$ are, respectively, the fundamental scale
and the bulk cosmological constant in the 6-dimensional world, whereas
$g_5$ and $\tilde g_5$
are the determinants of the induced metrics on the 5-dimensional hypersurfaces.

The solutions to the 6-dimensional Einstein's equations for negative 
bulk cosmological constant are given by 
\begin{equation}
a(y) =  e^{-c|y|} \hspace{1.5cm} 
b(z) =  \dis \frac{\cosh{(kz)}}{\cosh{(k\pi)}} \hspace{1.5cm} 
c  =  \dis \frac{R_y k}{r_z\cosh{k\pi}} \hspace{1.5cm}
k =  \dis r_z\sqrt{\frac{-\Lambda}{10 M_6^4}} \equiv r_z \,k' \ .
\label{eq:RS6_eqns}
\end{equation}
The junctures of the 4-branes constitute 3-branes. 
Both warp factors $a(y)$ and $b(z)$ are 
minimized at $(y,z)=(\pi,0)$ and we identify this 3-brane as the
one containing the SM. The resulting hierarchy factor is
\begin{equation}
w =  e^{-c \pi} \, {\rm sech}(k\pi) \, .
\label{eq:hierarchy}
\end{equation}
With the natural scale of the Higgs mass being given by $\Lambda_{\rm NP}$,
the cutoff scale for the SM, the observed Higgs mass is given by
\begin{equation}
m_H \quad = w \, \Lambda_{\rm NP} =  w \, \zeta \, {\rm min}(R_y^{-1},r_z^{-1}) \,\, ,
\label{eq:w_and_mh}
\end{equation}
where $1 \lapp \zeta \lapp 10 $ parametrizes the uncertainty in 
$\Lambda_{\rm NP}$, which must be smaller than $M_6$.
In order to explain a large hierarchy without introducing an 
unnatural separation of scales between the moduli, Eq.\ref{eq:hierarchy} along with 
the relation between $c$ and $k$ in Eq.\ref{eq:RS6_eqns} mandates that 
the warping in one of the two extra dimensions be substantially 
larger than the other.
This can be achieved by having 
($i$) a large ($\sim 10$) value for $k$ accompanied by an infinitesimally 
small $c$,  or,  
($ii$) a large ($\sim 10$) value for $c$ with a moderately small $k$.

%%%%%%%%%%%%%%%%%%%%%%%%%%%%%%%%%%%%%%%%%%%%%%%%%%%%%%%%%%%%%%%%%%%%%%%
% End of The Model
%%%%%%%%%%%%%%%%%%%%%%%%%%%%%%%%%%%%%%%%%%%%%%%%%%%%%%%%%%%%%%%%%%%%%%%

%%%%%%%%%%%%%%%%%%%%%%%%%%%%%%%%%%%%%%%%%%%%%%%%%%%%%%%%%%%%%%%%%%%%%%%
% KK modes
%%%%%%%%%%%%%%%%%%%%%%%%%%%%%%%%%%%%%%%%%%%%%%%%%%%%%%%%%%%%%%%%%%%%%%%
\section{The graviton KK modes}
\label{sec:KKmodes}
The dynamics of the 
fluctuations about the aforementioned classical 
metric configuration can be derived within a 
semi-classical approach\footnote{Note that using such an approach one can 
only perform tree-level calculations; evaluation of amplitudes involving 
graviton loops is forbidden. However, loop-diagrams (QCD or electroweak) 
where the graviton appears only in the tree level sub-diagrams, are 
permissible.}~\cite{Arun:2014dga}. 
Subsequent imposition of Neumann boundary conditions leads to the quantization
of graviton masses.
The consequent spectrum is qualitatively
different for the two distinct regions of the parameter space, namely,
large $k$ and small $k$. In the small $k$ domain, the graviton KK
modes, as expected, form nested towers characterised by the winding numbers 
$n$ and $p$. On the other hand, in the large $k$ domain, the mass difference
between successive $n$-levels is so large that only the $n$=0 mode
turns out to be relevant in the context of current experiments.
The interaction of a graviton with the SM field localized on the 3-brane,  
is described by 
\begin{equation} 
{\cal L}_{\rm int}= 
T^{\mu\nu}\Big(C_{00} h_{\mu\nu}^{(0,0)}+
\sum_{n\neq0}C_{n0} h_{\mu\nu}^{(n,0)}+
\sum _{n,p\neq0}C_{np} h_{\mu\nu}^{(n,p)} \Big) \,
\label{eq:Lint}
\end{equation} 
with $T^{\mu \nu}$ being the energy-momentum tensor of the corresponding field.
While the detailed results can be found in Ref.~\cite{Arun:2014dga}, 
we list the important formulae below.  

%%%%%%%%%%%%%%%%%%%%%%%%%%%%%%%%%%%%%%%%%%%%%%%%%%%%%%%%%%%%%
% Mass and Couplings - large k
%%%%%%%%%%%%%%%%%%%%%%%%%%%%%%%%%%%%%%%%%%%%%%%%%%%%%%%%%%%%%
\subsection{\texorpdfstring{Mass spectrum and couplings for the KK graviton 
                            in large $k$ regime}
                           {Mass spectrum and couplings for the KK graviton 
                            in large k regime}
           }
\label{sec:large_k_analytic}

When $k$ is large and $R_y$ \& $r_z$ have comparable magnitudes,
there is little or no warping in the $y$-direction. Hence, to the lowest order, the $y$-momentum eigenstates are 
just plane waves. This yields $m_{np}^2 \approx m_p^2 + n^2 R_y^{-2}$, and modes 
corresponding to $n \neq 0$ are too heavy to be of any consequence to LHC 
experiments. Effectively, we are left with a single tower of gravitons with 
masses $m_{0p} \approx m_p$. For $p=0$, $m_p = 0$. For $p \neq 0$, $\nu_p = 2p + 1/2$ and $m_p$ is obtained using the relation
\begin{equation}
\barr{rcl}
\nu_p & \equiv & \dis 
       \sqrt{4 + \frac{m_p^2 \, R_y^2}{c^2} } \, - \, \frac{1}{2} \ .
\earr
\label{eq:nup_def}
\end{equation}

Up to leading order in $c$, the couplings are given by : 
\begin{equation}
C_{00}  = 
\frac{\cosh^{3/2}(k\pi)}{M_6^{2}\sqrt{2 \pi \, R_y \, r_z \, B_0}} \ ,  
\qquad 
C_{0p} = 
\frac{\cosh^{3/2}(k \pi)}{M_6^{2}\sqrt{2 \pi \, R_y \, r_z \, B_p}} \,  
Q_{\nu_p}^{5/2}\left(0 \right) \ ,
\label{eq:coup_largek}
\end{equation}
where
\begin{equation}
B_{0} = \int_{-\pi}^{\pi} \cosh^3(k \, z)\,  dz \ ,
\qquad
B_{p \, \neq \, 0} = \int_{-\pi}^{\pi} {\rm sech}^2(k \, z) 
\left[Q_{\nu_p}^{5/2}(\tanh(k \, z)) \right]^2 dz \ .
\label{eq:Bp_largek}
\end{equation}
%
%%%%%%%%%%%%%%%%%%%%%%%%%%%%%%%%%%%%%%%%%%%%%%%%%%%%%%%%%%%%%
% End of Mass and Couplings - large k
%%%%%%%%%%%%%%%%%%%%%%%%%%%%%%%%%%%%%%%%%%%%%%%%%%%%%%%%%%%%%

%%%%%%%%%%%%%%%%%%%%%%%%%%%%%%%%%%%%%%%%%%%%%%%%%%%%%%%%%%%%%
% Mass and Couplings - small k
%%%%%%%%%%%%%%%%%%%%%%%%%%%%%%%%%%%%%%%%%%%%%%%%%%%%%%%%%%%%%

\subsection{\texorpdfstring{Mass spectrum and couplings for the KK graviton 
                            in small $k$ regime}
                           {Mass spectrum and couplings for the KK graviton 
                            in small k regime}
           }
\label{sec:small_k_analytic}

Small $k$ implies that the warping in the $z$-direction is small 
(though not as insignificant as the $y$-warping in the previous 
case), the mass scale in this direction is substantially high. 
In other words, $m_{10} \ll m_{01}$. Once again, $\nu_p = 3/2$ for $p=0$.
$m_{np}$ is obtained from 
\begin{equation}
m_{np} = \frac{x_{np} c}{e^{- c \pi} R_y}  \ .
\label{eq:cotthetap}
\end{equation}
where $x_{np}$ are solutions of 
\begin{equation}
\barr{rcl}
\dis 2 x_{np} J_{\nu_p -\frac{1}{2}}(x_{np} ) +(3-2 \nu_p )
J_{\nu_p +\frac{1}{2}}(x_{np} ) \, = \, 0 \ ,
\earr
\label{eq:xnpsoln}
\end{equation}
$J_{\alpha}$'s being Bessel functions of the first kind
(see Ref.~\cite{Arun:2014dga} for details). 
The couplings are given by
\begin{equation}
\barr{rcl}
C_{n0} 
& = & \dis
  \frac{e^{c \, \pi}}{M_6^{2}r_z} \, \cosh(k \, \pi)\, 
      \sqrt{\frac{k}{2 \, A_{n0} \, B_0}} \,  J_{2}(\theta_\pi) 
\\[2ex]
C_{n,{p\neq0}} &= & \dis
 \frac{e^{c \, \pi}}{M_6^{2}r_z} \,  \cosh(k \, \pi)\, 
  \sqrt{\frac{k}{2 \, A_{np} \, B_p}} \,
            J_{\nu_p + \frac{1}{2}}(\theta_\pi) \, 
	    \left[ \cot\theta_p \, P_{\nu_p}^{5/2}(0)
           + \, Q_{\nu_p}^{5/2}(0) \right] \ ,
\earr
\label{eq:coup_largec}
\end{equation}
where $B_{p = 0}$ is as before and 
\begin{equation}
A_{np} = \int_{0}^{1} r \, 
\left[J_{\nu_p +\frac{1}{2}}(x_{np} \, r)\right]^2 dr \ ,
\qquad
B_{p \, \neq \, 0} = \int_{-\tau_\pi}^{\tau_\pi} dr \,
    \left[ \cot\theta_p \, P_{\nu_p}^{5/2}(r)
           + \, Q_{\nu_p}^{5/2}(r) \right]^2  \ .
\label{eq:Anp_largec}
\end{equation}
           
%%%%%%%%%%%%%%%%%%%%%%%%%%%%%%%%%%%%%%%%%%%%%%%%%%%%%%%%%%%%%
% End of Mass and Couplings - small k
%%%%%%%%%%%%%%%%%%%%%%%%%%%%%%%%%%%%%%%%%%%%%%%%%%%%%%%%%%%%%

%%%%%%%%%%%%%%%%%%%%%%%%%%%%%%%%%%%%%%%%%%%%%%%%%%%%%%%%%%%%%%%%%%%%%%%
% Constraining the Parameter Space
%%%%%%%%%%%%%%%%%%%%%%%%%%%%%%%%%%%%%%%%%%%%%%%%%%%%%%%%%%%%%%%%%%%%%%%
\section{Constraining the Parameter Space}
\label{sec:limits}

In this section we examine the viability of the model described in the 
previous section in the light of the present LHC data. Furthermore, we 
confront it with existing projections for Run 2. Before proceeding further, 
we define two more (dimensionless) quantities :
\begin{equation}
\epsilon \quad \equiv \quad \dfrac{k}{r_zM_6} \quad = \sqrt{\frac{-\Lambda}{10 M_6^6}} \qquad , \qquad
\alpha \quad \equiv \quad \dfrac{R_y}{r_z}
\end{equation}
$\epsilon$ is related to the bulk curvature and the validity of the 
semi-classical approximations used in extremizing the Einstein-Hilbert action 
requires $\epsilon < 0.1$. Furthermore, we do not expect $R_y$ and $r_z$ to be 
vastly different in magnitude since that would lead to a hierarchy in
the moduli. Therefore we will only consider $10^{-3} < \alpha < 10^3$.

\begin{figure}[!htbp] 
\subfigure[]{\includegraphics[scale=0.27]{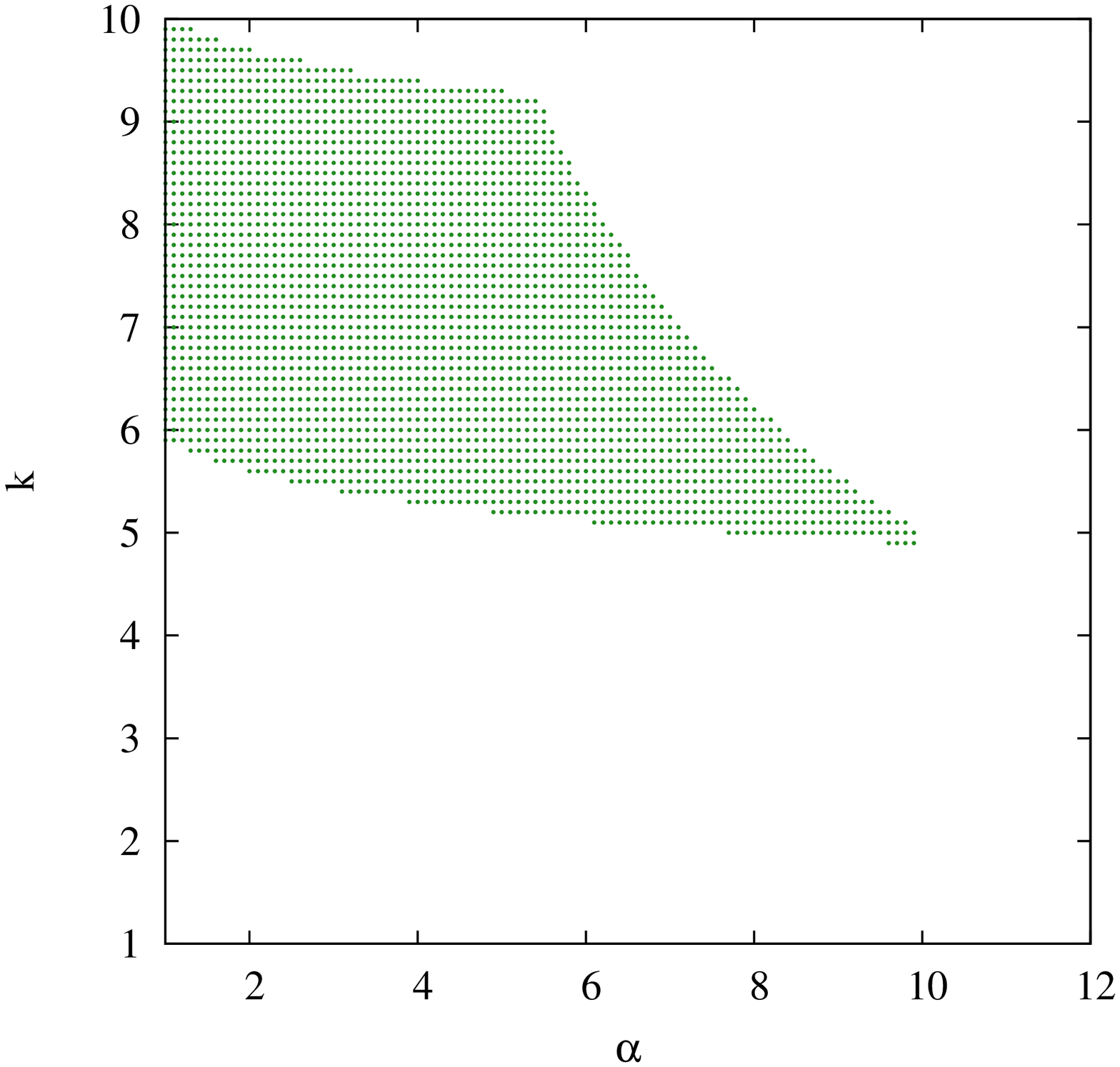}}
\subfigure[]{\includegraphics[scale=0.27]{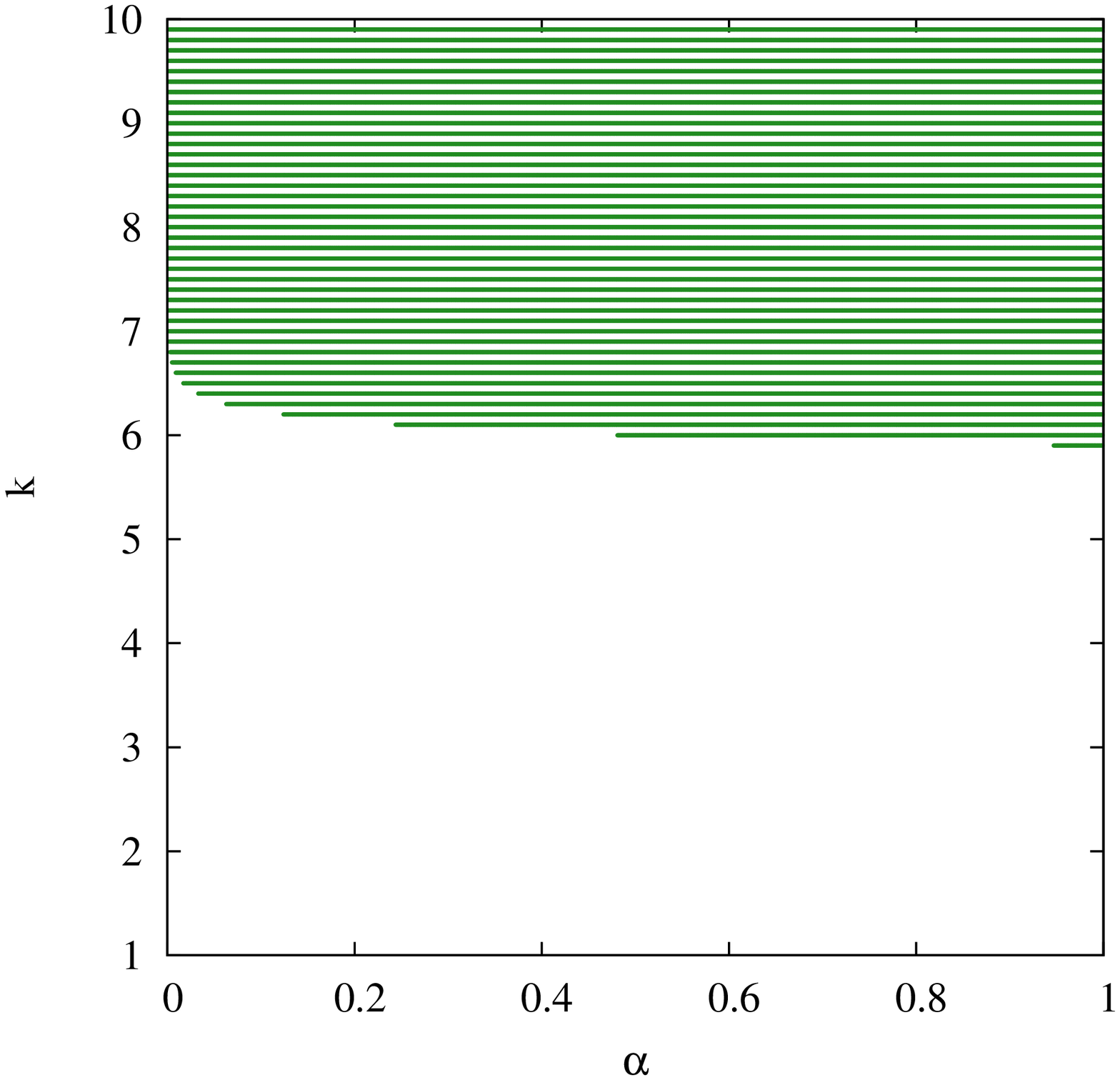}}
\subfigure[]{\includegraphics[scale=0.27]{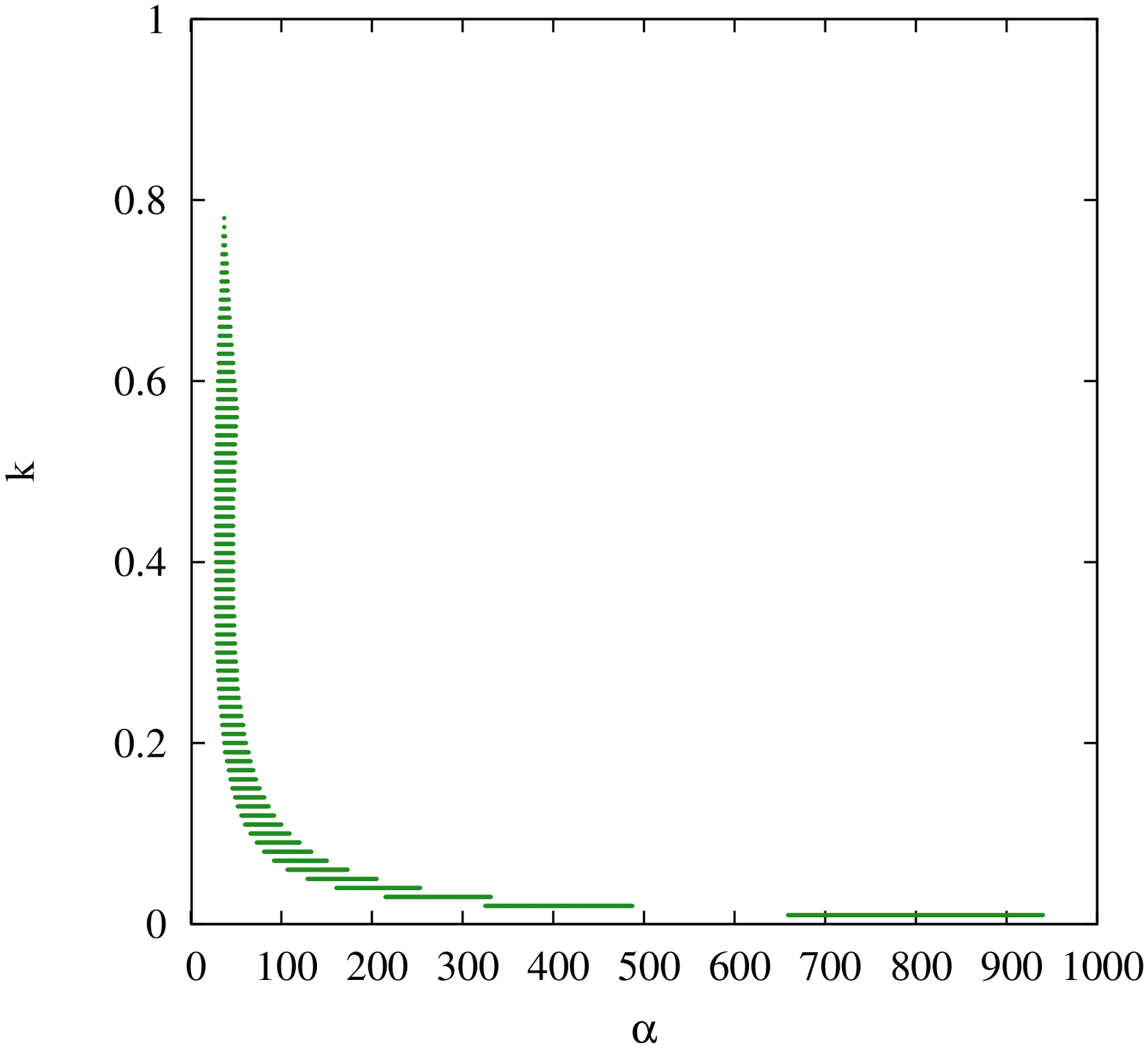}}
\caption{The relevant region of parameter space for $\zeta$ = 1 and 
(a) large $k$, large $\alpha$,
(b) large $k$, small $\alpha$, and  
(c) small $k$, large $\alpha$. 
For small $k$, small $\alpha$, the entire region is ruled out by the 
constraints discussed above.}
\label{fig:basic}
\end{figure}

In order to perform a systematic scan, we divide the parameter space into 
the following 4 regions :
\vspace*{-10pt}
\begin{itemize}
 \item large $k$, large $\alpha$ : $1 \leqslant k \leqslant 10$ ; 
                                   $1 \leqslant \alpha \leqslant 10^3$
 \vspace*{-8pt}                                    
 \item large $k$, small $\alpha$ : $1 \leqslant k \leqslant 10$ ; 
                                   $10^{-3} \leqslant \alpha < 1$
 \vspace*{-8pt}                                    
 \item small $k$, large $\alpha$ : $0.01 \leqslant k < 1$ ; 
                                   $1 \leqslant \alpha \leqslant 10^3$
\end{itemize}
\vspace*{-10pt}
In each regime, we calculate the mass of the 
lightest graviton ($m_{01}$ for large $k$ and $m_{10}$ for small $k$),
holding $m_H = 125 \gev$ and restricting 
$0.0001 \leqslant \epsilon \leqslant 0.1$ and $M_6 < M_P$. 
The ensuing part of the parameter space that supports a KK-graviton 
within the LHC reach is depicted in Fig.\ref{fig:basic}, 
and for the remainder of this paper, we consider only these subsets. 
Note the region small $k$, small $\alpha$
gets completely ruled out as the warping in this domain is not large enough to reproduce
the hierarchy between Planck scale and TeV scale. 

%%%%%%%%%%%%%%%%%%%%%%%%%%%%%%%%%%%%%%%%%%%%%%%%%%%%%%%%%%%%%
% ATLAS Limits
%%%%%%%%%%%%%%%%%%%%%%%%%%%%%%%%%%%%%%%%%%%%%%%%%%%%%%%%%%%%%
\subsection{\texorpdfstring{Data from the LHC at $\sqrt{s}$ = 8 TeV}
                           {Data from the LHC at 8 TeV}
           }
\label{sec:ATLAS_limits}

Diphoton production is often the preferred channel for gravition searches
as it provides a clean signature. The experimental mass resolution in this channel 
is similar to that for leptons but the branching fraction is larger ($BR (G \to \gamma\gamma)$ = 2\,$BR(G \to \ell^+ \ell^-)$).
The ATLAS Collaboration conducted a search for high invariant mass diphoton pairs resulting from RS graviton decays in the 
8 TeV Run of the LHC~\cite{ATLAS2015}. They found the data to be consistent with the SM. 
To interpret the consequent limits in the present context,  we must first establish the correspondence 
between the parameters of the two models. For the RS model, the mass of the lighest KK 
graviton is given by
\begin{equation}
m_1 \, = \, m_{G^*} \, = \, x_1 \, k_5e^{\pi k_5 R_c} \ ,
\label{eq:m_5D}
\end{equation}
where $x_1$ is the first root of the Bessel function $J_1(x)$.
The corresponding graviton interaction term is given by 
\begin{equation} 
L_{\rm int}^{\rm 5D RS} 
\quad = \quad 
\dfrac{k_5/\mpl}{k_5e^{\pi k_5 R_c}} \,  
\sum_{n=1}^{\infty} T^{\mu\nu}(x) \, h_{\mu\nu}^{(n)} (x) 
\quad = \quad 
\dfrac{k_5/\mpl}{m_{G^*}/x_1} \,  
\sum_{n=1}^{\infty} T^{\mu\nu}(x) \, h_{\mu\nu}^{(n)} (x) \ .
\label{eq:Lint5D}
\end{equation} 
Comparing Eq.\ref{eq:Lint5D} with Eq.\ref{eq:Lint}, it 
is only natural to bridge the 5-dimensional and 6-dimensional models with
\begin{equation}
\dfrac{k_5/\mpl}{m_{G^*}/x_1}
\longleftrightarrow 
|C_{01}|, \, |C_{10}| \ ,
\qquad 
m_{G^*}
\longleftrightarrow 
|m_{01}|, \, |m_{10}| \ .
\label{eq:5D6D}
\end{equation}
as they form equivalent descriptions for the collider analysis.  With
this mapping in place, the remainder of the analysis follows exactly
as that for the 5-dimensional RS model. For a graviton of a given mass
and a given final state (diphoton in this case), the phase-space
distribution of the final state particles is identical 
for the two models.  Accordingly, detector
resolution, efficiency of cuts and overall experimental sensitivity
would also be practically identical.  On the theoretical
front, QCD and electroweak NLO corrections too would be identical to those
for the RS model (see Footnote 1). Hence one can directly identify the
ATLAS limits on the $k_5/\mpl - m_{G^*}$ plane onto the
$C_{01} - m_{01}$ \Big($C_{10} - m_{10}$\Big ) plane for the large $k$
\Big(small $k$\Big) case.
\begin{figure}[!ht]
\centering
\subfigure[]{\includegraphics[scale=0.35]{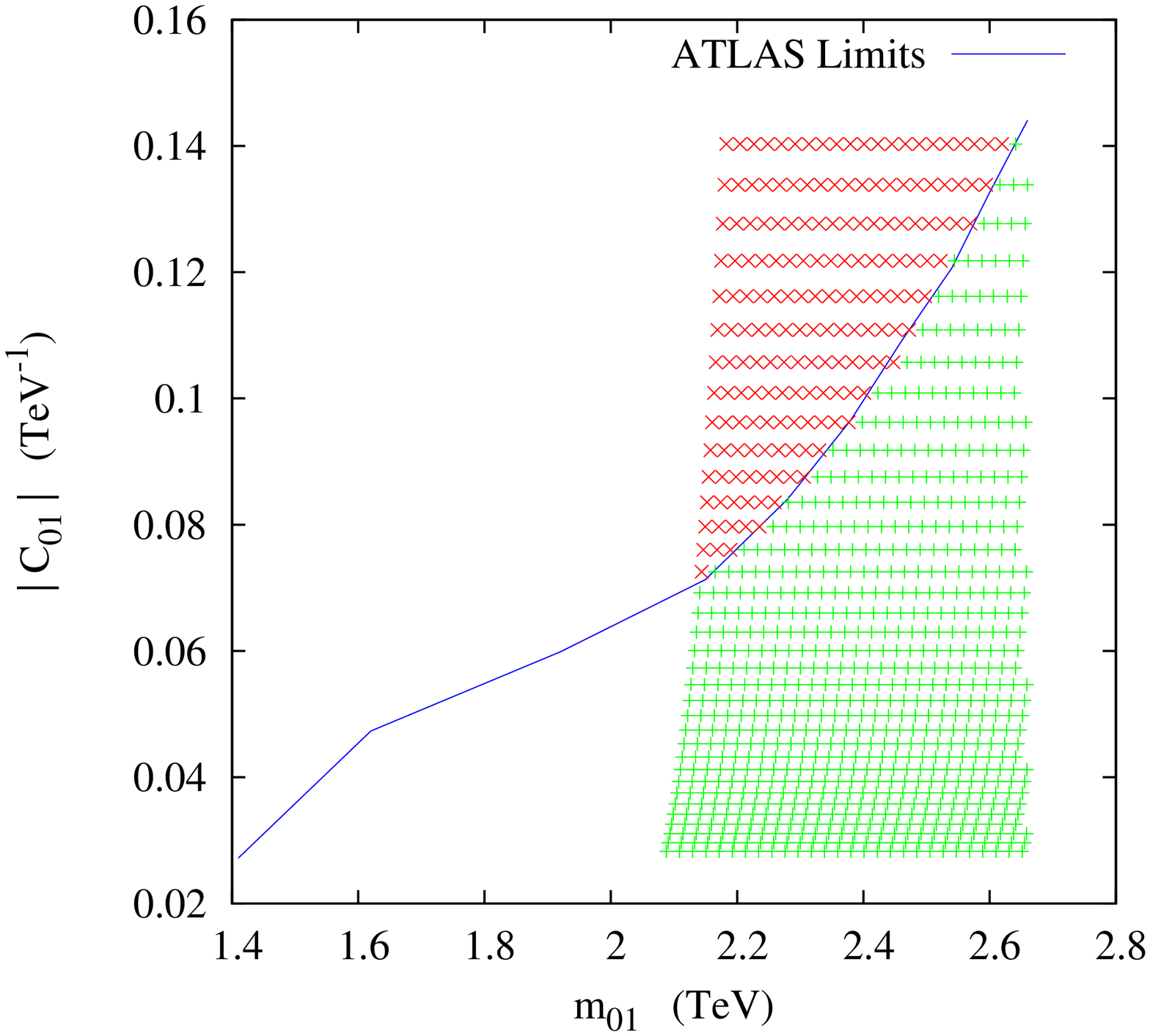}} 
\subfigure[]{\includegraphics[scale=0.35]{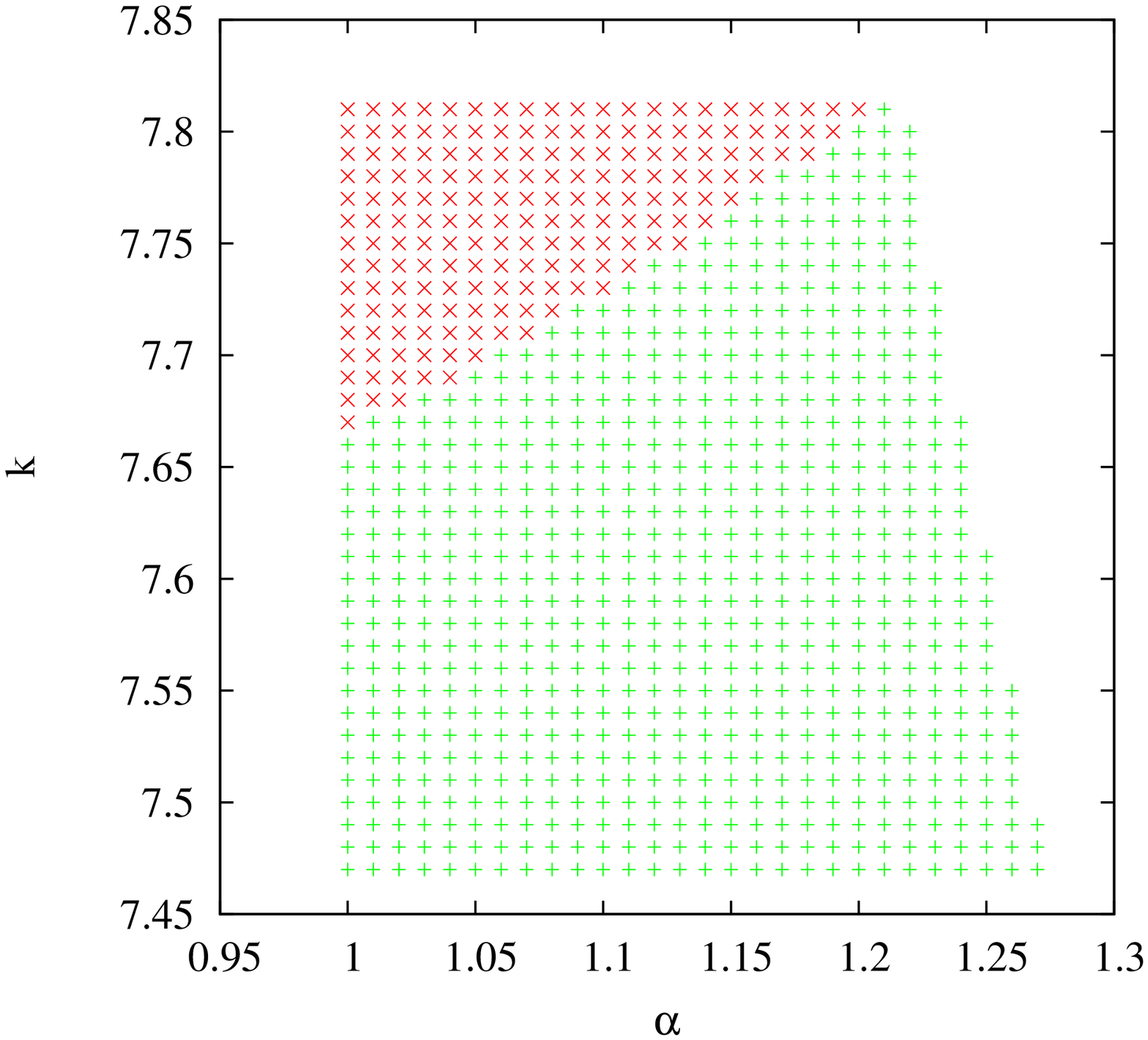}}
\caption{The red $\times$'s (green $+$'s) show the parts of large $k$, large $\alpha$ 
parameter space (for $\zeta$ = 1) ruled out (allowed) by the ATLAS\cite{ATLAS2015} limits 
(blue curve). All the points depicted satisfy the initial requirements on $\epsilon$ and $M_6$.}
\label{fig:ATLAS_largeK_largeA}
\end{figure}

Fig.\ref{fig:ATLAS_largeK_largeA}(a) shows the limits on the 
$C_{01} - m_{01}$ plane for the large $k$, large $\alpha$ scenario.  
Since the limits are based on 8 $\tev$ data, the sensitivity to KK graviton 
masses extends only upto 2.5 $-$ 3 $\tev$. The $\times$'s above the curve are 
show the region that is ruled out. The corresponding region in the 
$k - \alpha$ plane is shown in Fig.\ref{fig:ATLAS_largeK_largeA}(b). 
Once again $\times$'s denote the region that is ruled out. 

\begin{figure}[!ht]
\centering
\includegraphics[scale=0.35]{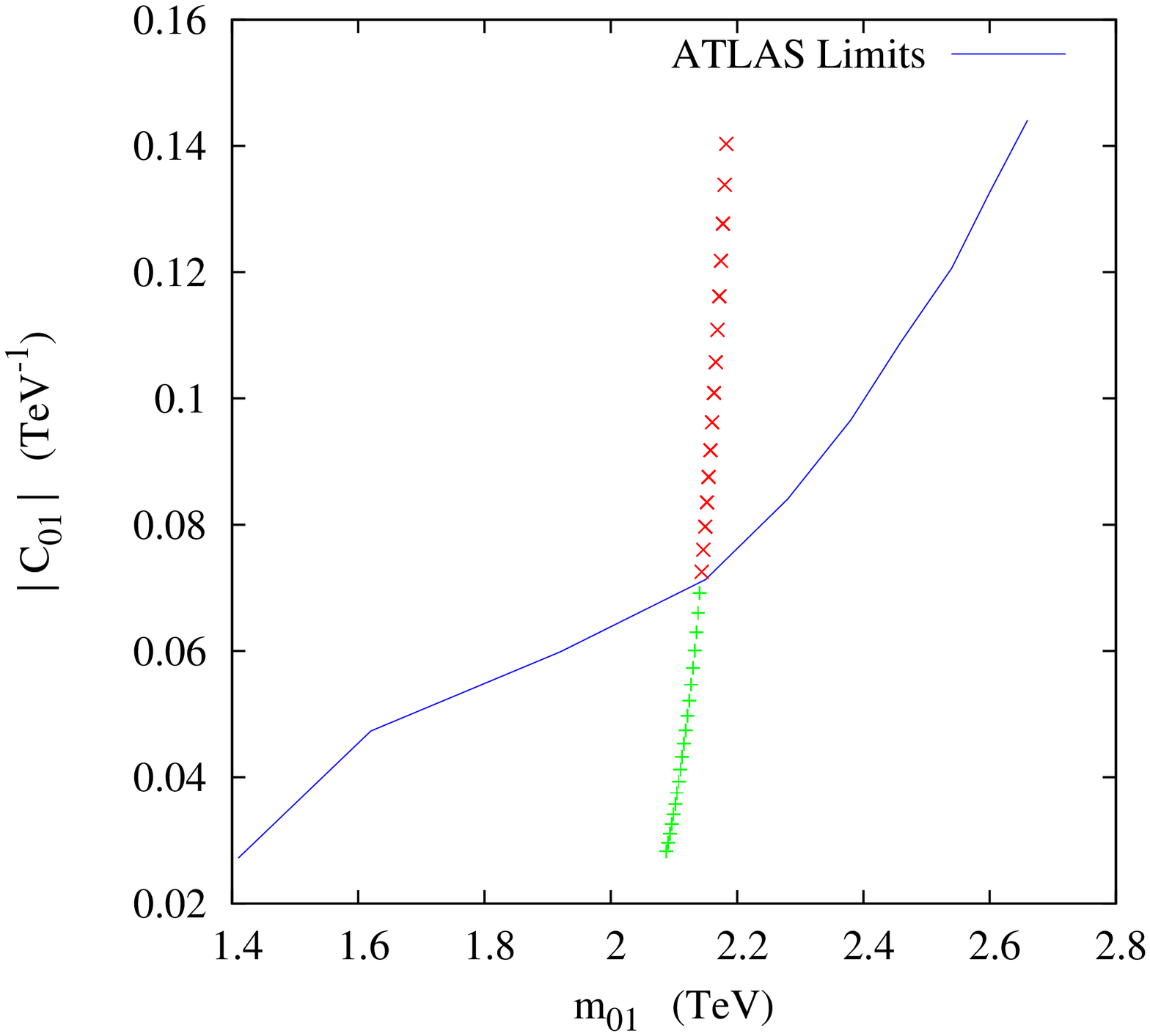}
\includegraphics[scale=0.35]{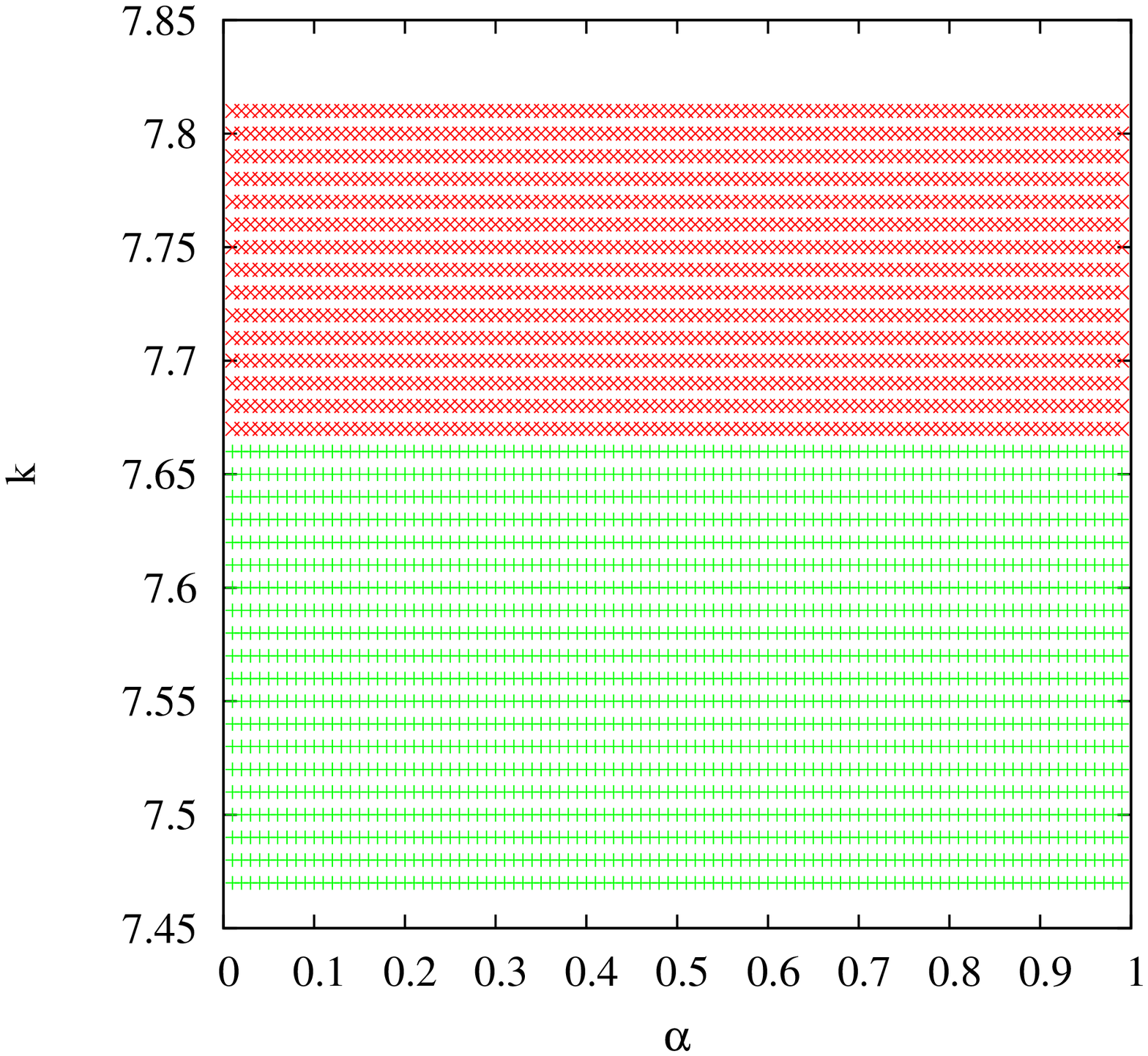}
\vskip -20pt
\caption{As in Fig.\ref{fig:ATLAS_largeK_largeA}, 
but for the large $k$, small $\alpha$ regime instead.}
\label{fig:ATLAS_largeK_smallA}
\end{figure}

A similar exercise can be carried out for the large $k$, small $\alpha$ case.
The results are depicted in Fig.\ref{fig:ATLAS_largeK_smallA}. 
On the other hand, the small $k$, large $\alpha$ region remains unconstrained 
by the 8 TeV LHC data as the values of $m_{10}$ are typically larger than 
3 TeV in this case.

%%%%%%%%%%%%%%%%%%%%%%%%%%%%%%%%%%%%%%%%%%%%%%%%%%%%%%%%%%%%%
% End of ATLAS Limits
%%%%%%%%%%%%%%%%%%%%%%%%%%%%%%%%%%%%%%%%%%%%%%%%%%%%%%%%%%%%%

%%%%%%%%%%%%%%%%%%%%%%%%%%%%%%%%%%%%%%%%%%%%%%%%%%%%%%%%%%%%%
% 14 TeV Projections
%%%%%%%%%%%%%%%%%%%%%%%%%%%%%%%%%%%%%%%%%%%%%%%%%%%%%%%%%%%%%
\subsection{14 TeV Projections}
\label{sec:14TeVProjections}

In Ref.~\cite{Das:2014tva}, the authors used Monte Carlo simulations at 
NLO along with parton showers, and obtained 
projections for the lower limits on $m_{G^*}$ that may be extracted from 
the $\ell^+\ell^-$ (Drell-Yan) and $\gamma\gamma$ (Diphoton) final states 
with an integrated luminosity of 50 fb$^{-1}$ for certain benchmark values of $k_5/\mpl$.
The results are presented in Table 1 and Table 2, respectively, of Ref.~\cite{Das:2014tva}.

%%%%%%%%%%%%%%%%%%%%%%%%%%%%%%%%%%%%%%%%%%%%%%%%%%%%%%%%%%%%%
% Figures-I
%%%%%%%%%%%%%%%%%%%%%%%%%%%%%%%%%%%%%%%%%%%%%%%%%%%%%%%%%%%%%
\begin{figure}[!ht]
\centering
\subfigure[]{\includegraphics[scale=0.35]{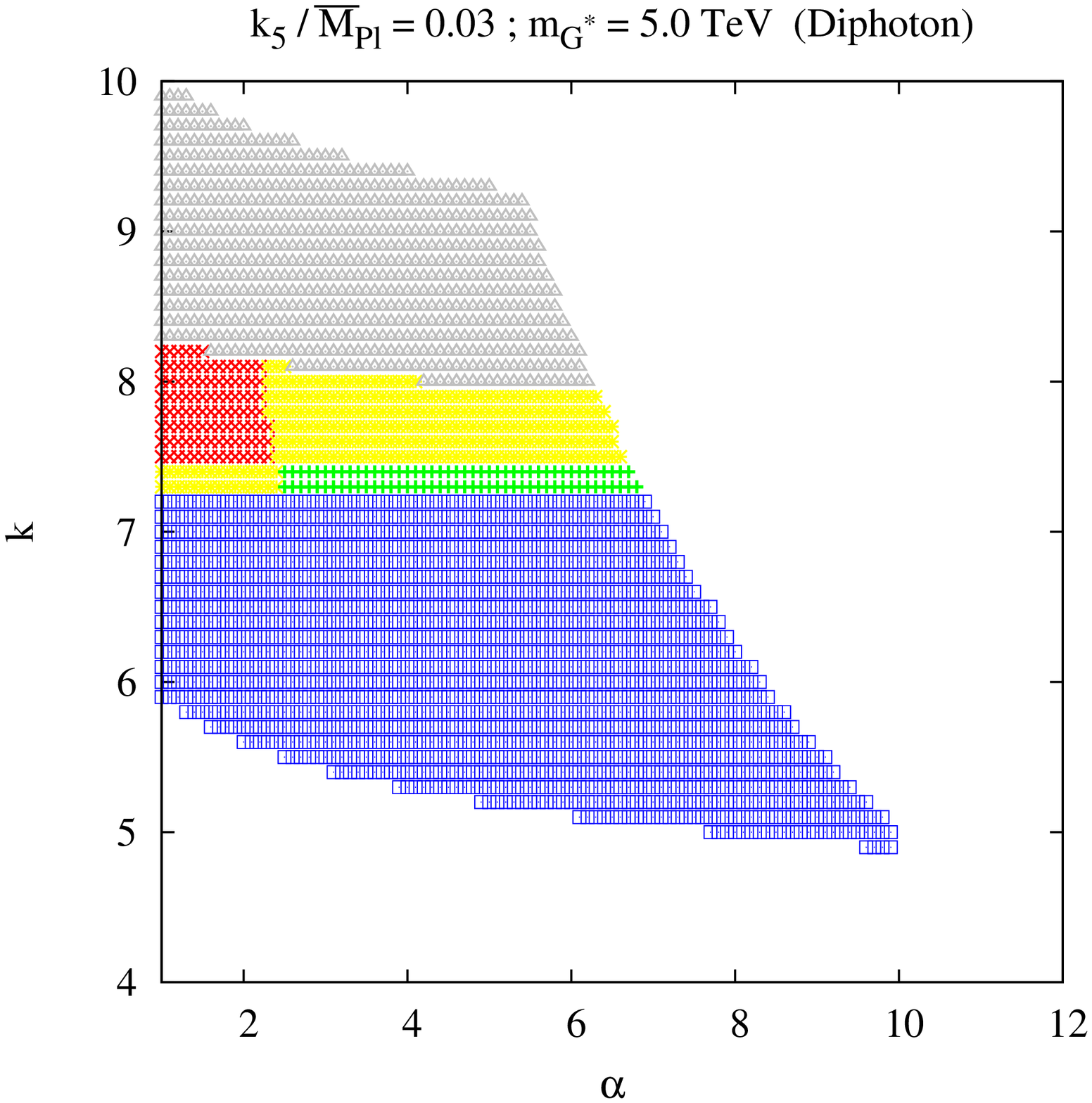}} \qquad \qquad 
\subfigure[]{\includegraphics[scale=0.35]{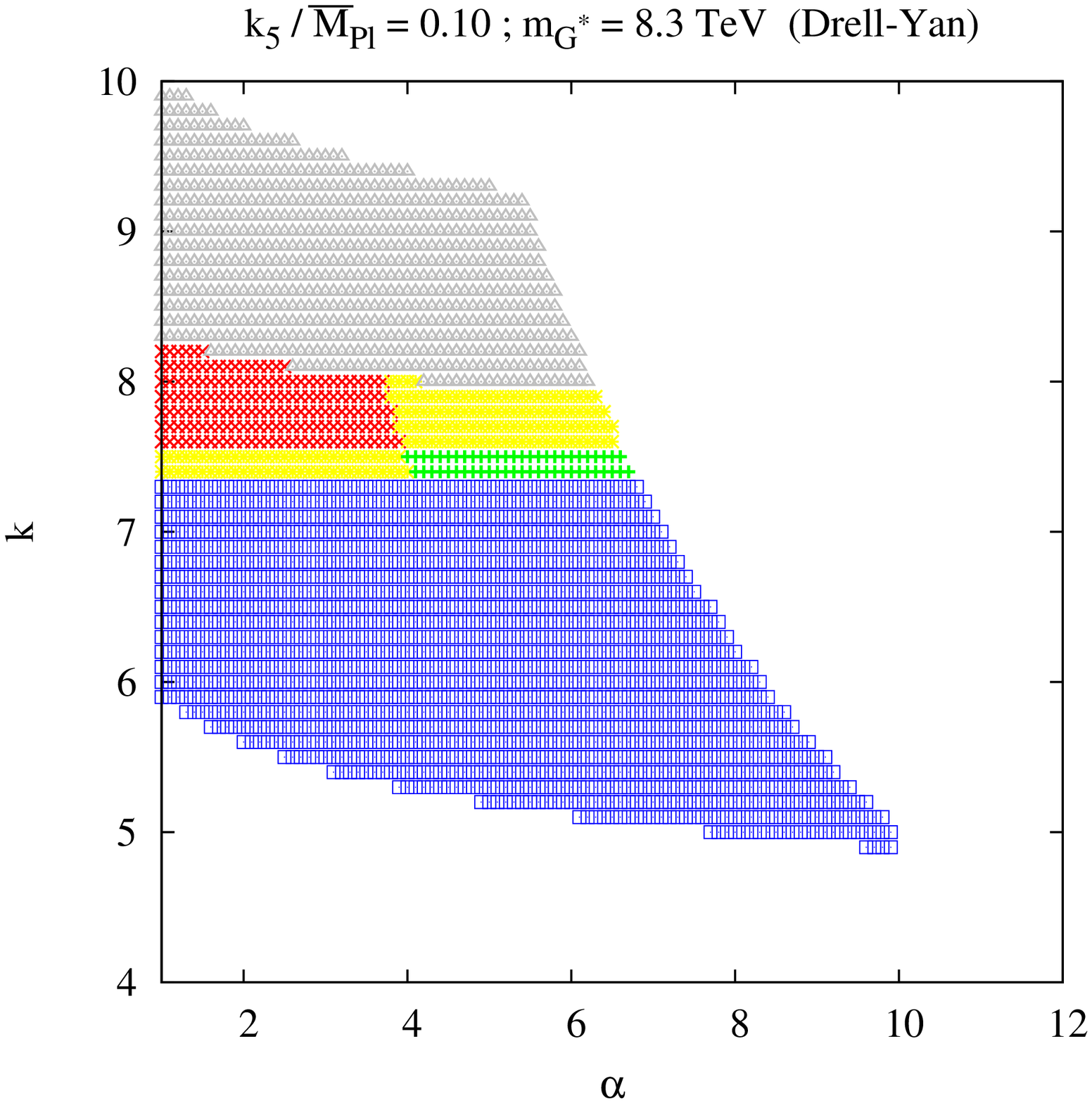}}

\subfigure[]{\includegraphics[scale=0.35]{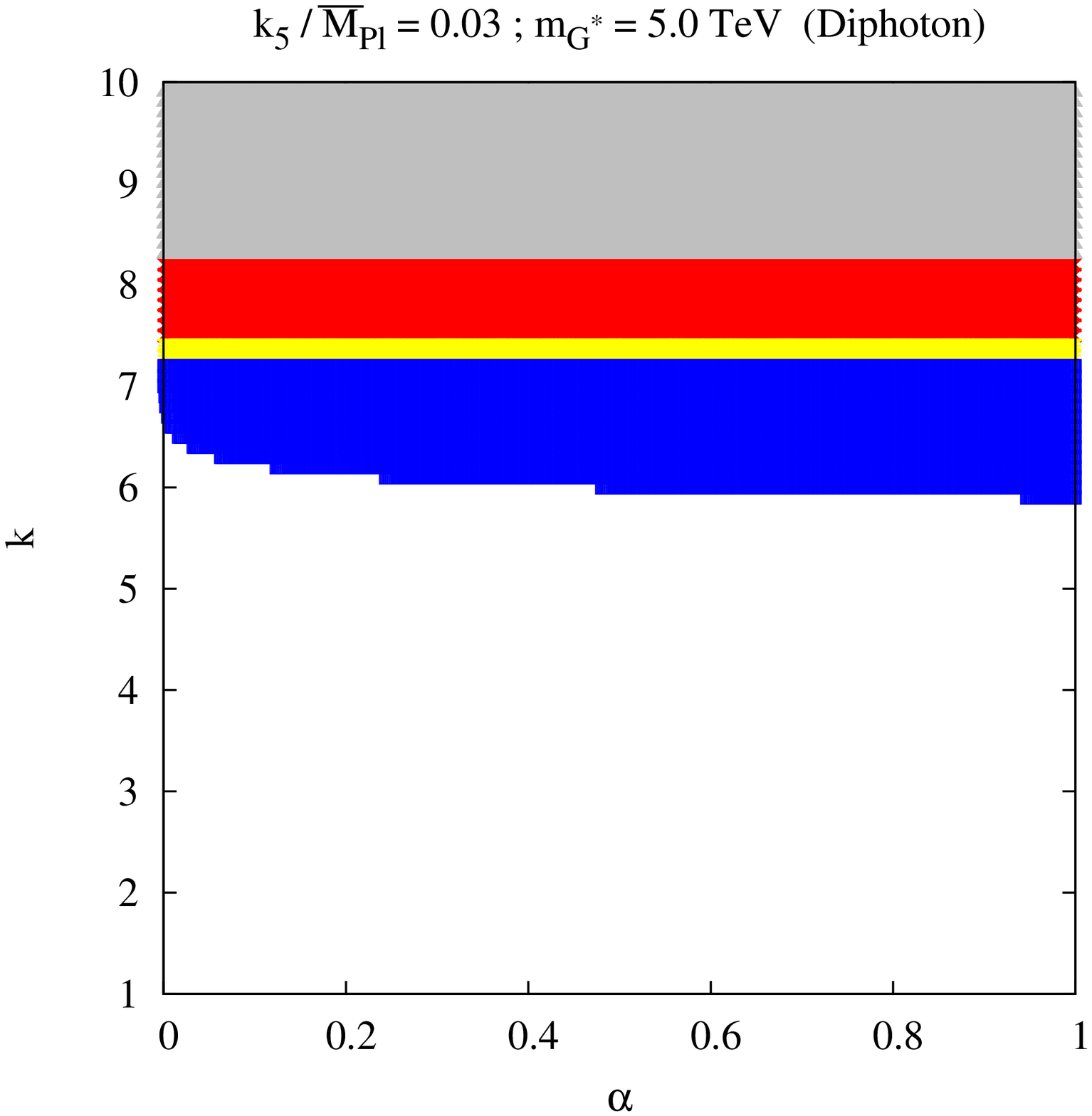}} \qquad \qquad 
\subfigure[]{\includegraphics[scale=0.35]{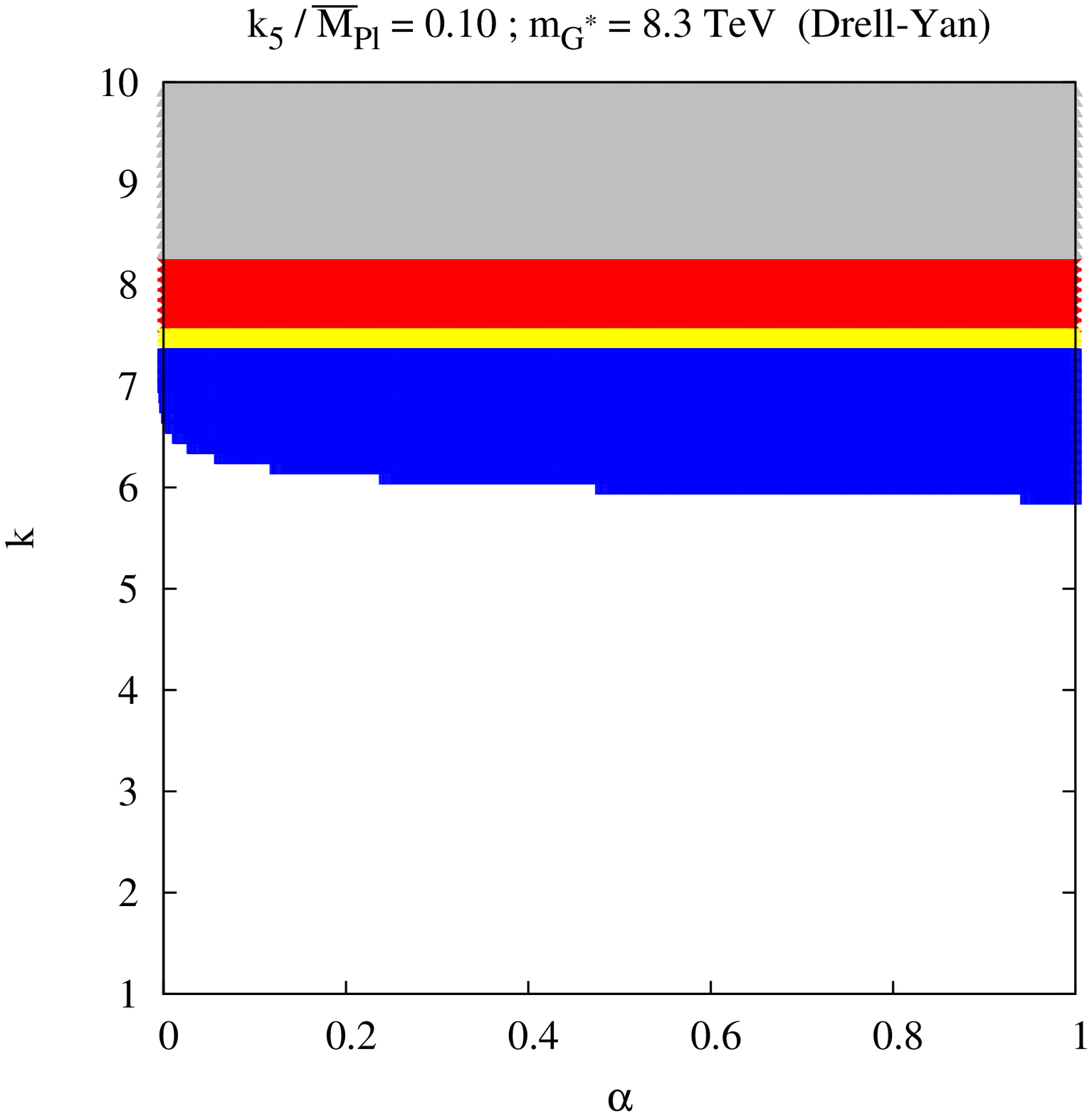}}

\caption{Exclusion of parameter space based on limits from 
Ref.\cite{Das:2014tva}. Upper row $-$ large $k$, large $\alpha$; lower row $-$ large $k$ small $\alpha$. $\zeta$ = 1. See text for detailed description.}
\label{fig:14TeV}
\end{figure}
%%%%%%%%%%%%%%%%%%%%%%%%%%%%%%%%%%%%%%%%%%%%%%%%%%%%%%%%%%%%%
% End of Figures-I
%%%%%%%%%%%%%%%%%%%%%%%%%%%%%%%%%%%%%%%%%%%%%%%%%%%%%%%%%%%%%

We now examine what these limits signify for the surviving parameter space of the 6-dimensional model
as depicted by Fig.\ref{fig:basic}.
At the outset, we note that if the coupling is very large, it can lead to the graviton's width being 
larger than its mass. Such couplings are clearly unphysical as they would 
invalidate any particle description for gravitons. The region of in the $k$-$\alpha$ plane 
that leads to such large, unphysical couplings is marked by grey triangles in Fig.~\ref{fig:14TeV}. 
The limits obtained in Ref.~\cite{Das:2014tva} are based on an integrated luminosity of 50 fb$^{-1}$.
The LHC is expected to accumulate about 3000 fb$^{-1}$ in its lifetime~\cite{SteveMyers}). 
For a given mass, the sensitivity to $C_{np}$ obtained from the above analysis would then be extended 
to $(50/3000)^{1/4} C_{np} \approx 0.36 C_{np}$. Values of $C_{np}$ smaller than this would not be 
probed by the LHC. The corresponding part of the $k$-$\alpha$ plane is marked by blue boxes in 
Fig.~\ref{fig:14TeV}. 

For $k_5/\mpl$ = 0.03, Ref.~\cite{Das:2014tva} finds the lower limit on $m_{G^*}$ to be 5.0 TeV.
This implies that for $C_{np} \geqslant$ 0.023, $m_{np} \leqslant$ 5.0 TeV would be ruled out. 
In Fig.~\ref{fig:14TeV}, this region is denoted in by red $\times$'s. The complementary region, with smaller couplings 
and larger masses is allowed and shown by green $+$'s. 
For smaller masses and couplings, the resonance may be observed with a lower significance, whereas 
if both the coupling and the mass are larger, the signal would take the form of a deviation in the tail of the invariant mass 
spectrum. Such regions are denoted, respectively, by lower-left and upper-right regions marked by yellow stars in 
Fig.~\ref{fig:14TeV}(a) \& (b) [large $k$, large $\alpha$]. 
In Fig.~\ref{fig:14TeV}(c) \& (d) [large $k$, small $\alpha$], the entire yellow region corresponds to smaller masses and couplings.
The small $k$, large $\alpha$ region leads to couplings that are too small to be probed by the LHC.
Hence the entire region in Fig~\ref{fig:basic}(c) would survive the LHC.

%%%%%%%%%%%%%%%%%%%%%%%%%%%%%%%%%%%%%%%%%%%%%%%%%%%%%%%%%%%%%
% End of 14 TeV Projections
%%%%%%%%%%%%%%%%%%%%%%%%%%%%%%%%%%%%%%%%%%%%%%%%%%%%%%%%%%%%%

%%%%%%%%%%%%%%%%%%%%%%%%%%%%%%%%%%%%%%%%%%%%%%%%%%%%%%%%%%%%%
% The 750 GeV excess
%%%%%%%%%%%%%%%%%%%%%%%%%%%%%%%%%%%%%%%%%%%%%%%%%%%%%%%%%%%%%
\section{\texorpdfstring{The excess at $m_{\gamma\gamma}$ = 750 GeV}
                           {The excess at 750 GeV}
           }
\label{sec:theexcess}

Recently, both the ATLAS~\cite{ATLASDec2015} and CMS~\cite{CMSDec2015} collaborations have 
reported an excess in the diphoton mass spectrum near $m_{\gamma\gamma} \sim$ 750 $\gev$. 
A spin-1 resonance interpretation for this excess is ruled out due to considerations of 
angular momentum conservation and the fact that the final state consists of identical particles\footnote{The spin-1 
interpretation would still be admissible if the photon pair were accompanied by a third, soft particle.}.
The ATLAS collaboration has analyzed the excess in the context of a Higgs-like (spin-0) particle. 
The CMS analysis has considered the RS-graviton interpretation and found the excess to be most compatible with 
$m_{G*}$ = 760 GeV for an effective coupling $k_5/\mpl$ = 0.01. In a later update~\cite{ATLASMar2016}, 
the ATLAS collaboration has presented a spin-2 analysis in which they find that the largest deviation from the 
background-only hypothesis occurs for signal hypothesis corresponding to $k_5/\mpl$ = 0.21 and $m_{G*}$ = 750 GeV.

The existence of an excess in the 13 $\tev$ data, when viewed in the
context of lack of any such excess in the 8 $\tev$ data points to 
gluon-gluon fusion as the dominant production mechanism. While many models have been proposed, 
most have sought to explain the excess in terms of a $J=0$ state. The CMS
analysis for the 5-dimensional RS scenario suggests that the observed rates are too low even for $k_5/\mpl = 0.01$,
a value already at the edge of the aesthetically acceptable region for $k_5/\mpl$. Indeed, this was
to be expected given the existing studies of RS gravitons. Furthermore,
with RS gravitons coupling universally to the SM fields, such a diphoton excess, would, 
be accompanied by similar excesses in other channels
(most notably in $e^+e^-, \mu^+\mu^-, W^+W^-, ZZ, t \bar t$ and $hh$), 
none of which have been seen.

As we have learned in the preceding section, the situation is markedly different
for 6-dimensional nested warping, on account of both the change in the spectrum as 
well as the coupling of the graviton to th SM fields. 
This opens up the possibility of a signal strength commensurate with the observed excess.
We now examine this in detail. The issue of the lack of excess in other channels remains and we will return to it 
at a later stage.

\begin{figure}[!ht]
\centering
\includegraphics[scale=0.27]{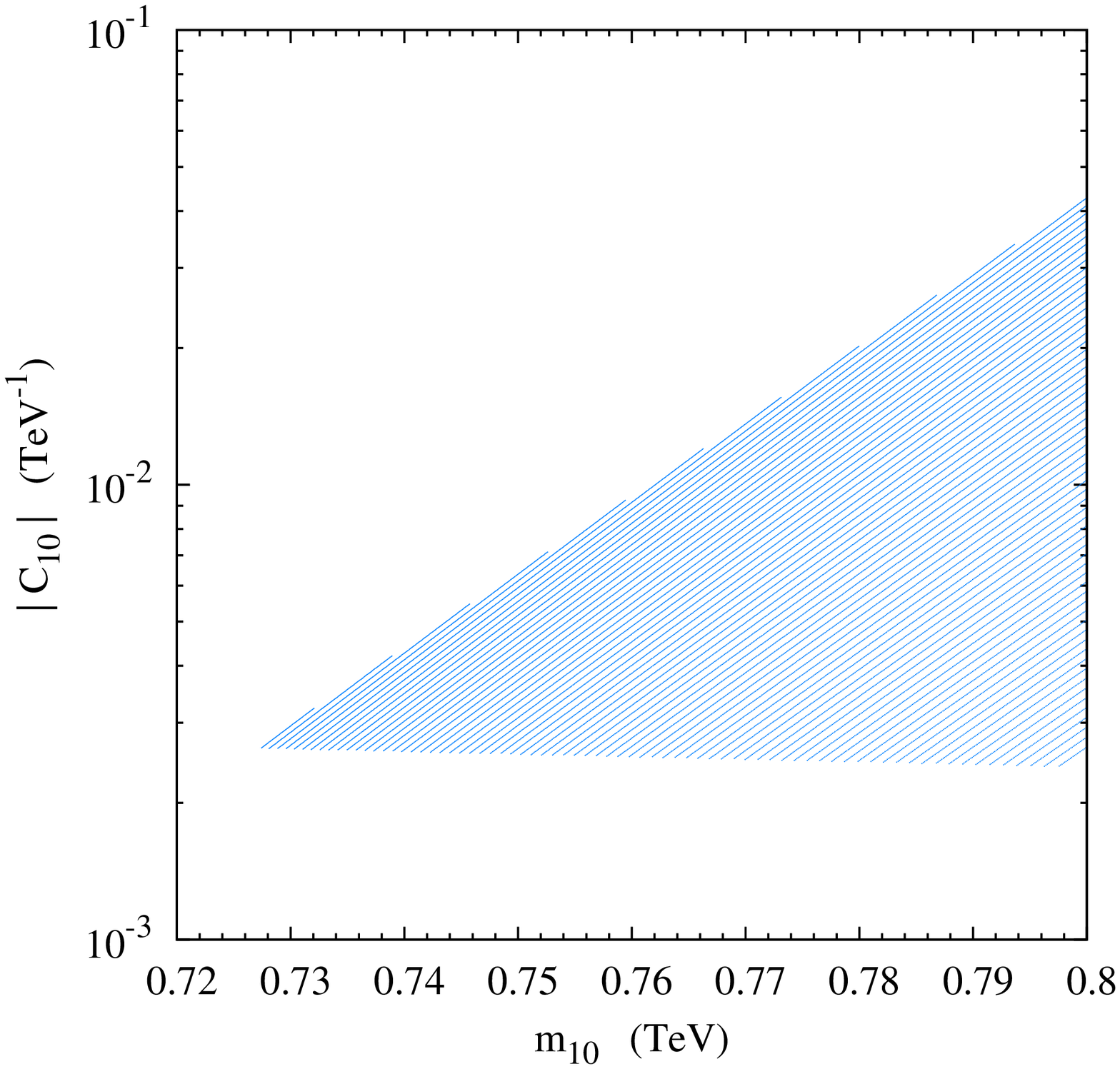}
\includegraphics[scale=0.27]{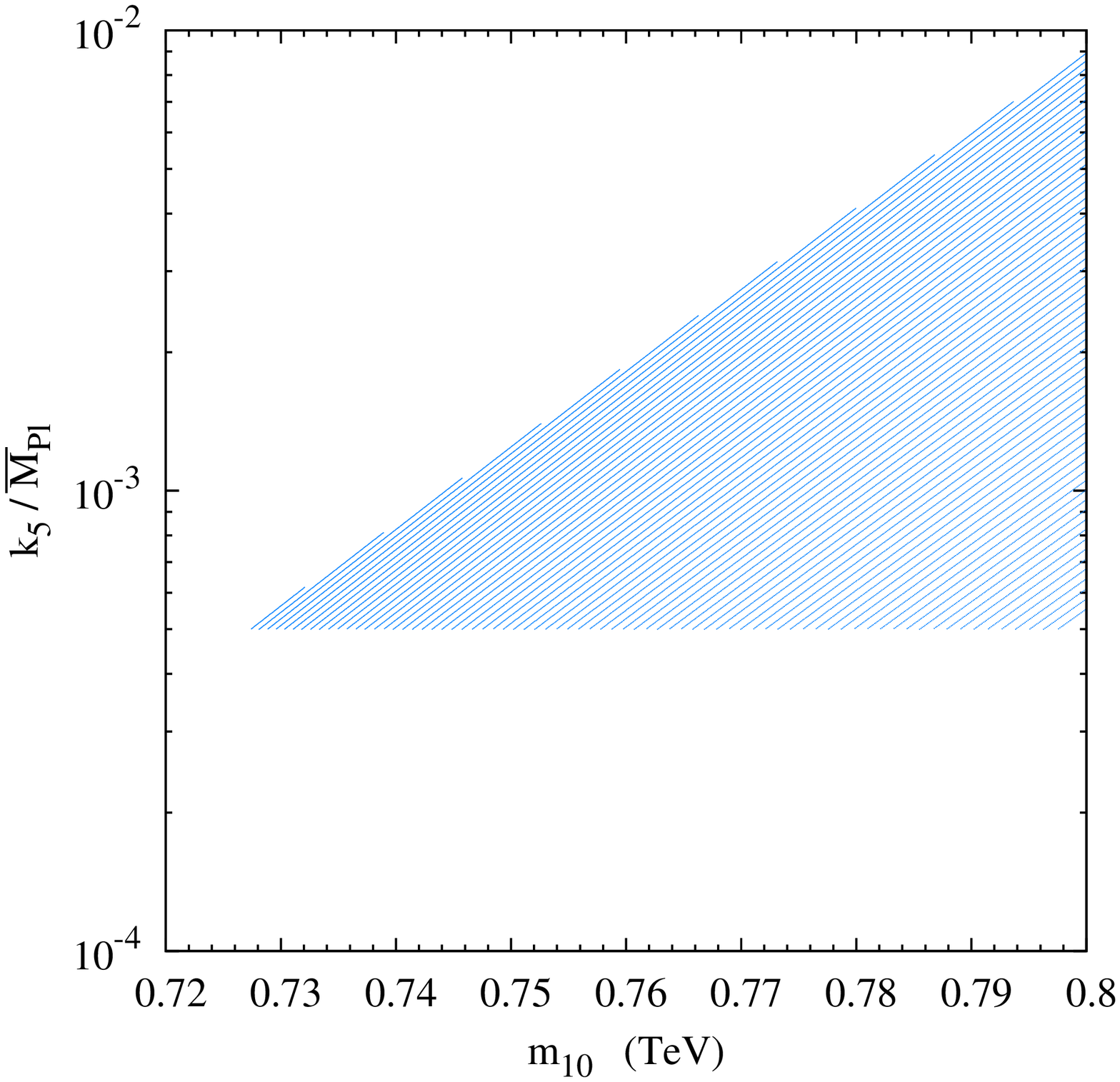}
\includegraphics[scale=0.28]{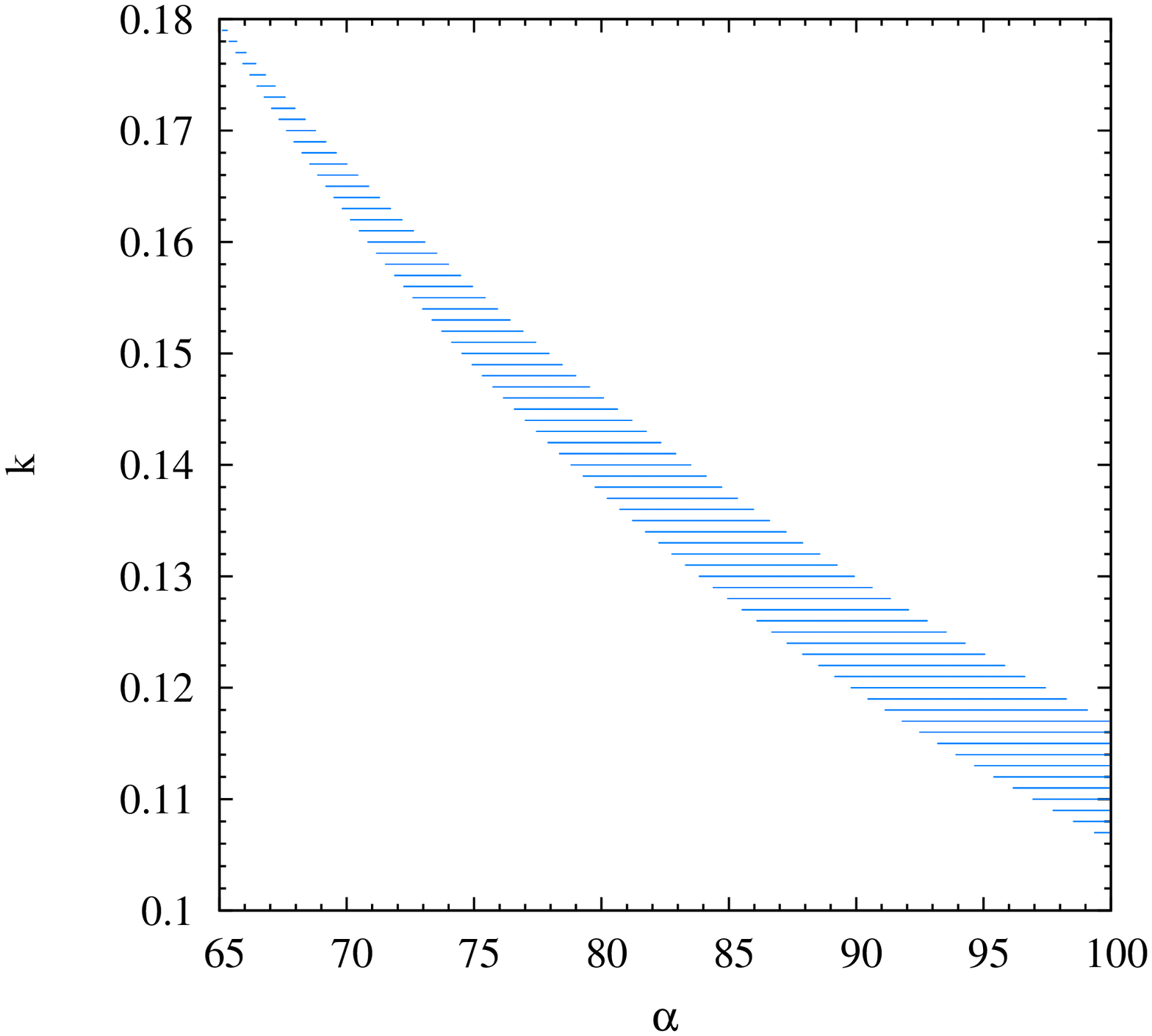}
\vskip -20pt
\caption{The region of parameter space, for small $k$, large $\alpha$ regime 
with $\Lambda_{\rm NP}=7 R_y^{-1}$, consistent with the recently reported 
excess in the diphoton invariant mass spectrum in the 13 TeV run of the 
LHC~\cite{ATLASDec2015,CMSDec2015}.}
\label{fig:750GeVsmallk}
\end{figure}

In the preceding sections, we have restricted ourselves to the case where 
$\zeta$ = 1, i.e. where $\Lambda_{\rm NP}$ for the SM is identified 
with $min(R_y^{-1},r_z^{-1})$. However, in order to have $m_{np} \in [700,800]$ $\gev$
along with suitable couplings, we need to allow $\zeta > 1$\footnote{We will, 
nonetheless restrict ourselves to $\zeta<10$ so as not to introduce a
new little hierarchy. For such values of $\zeta$, $min(R_y^{-1},r_z^{-1}) < \Lambda_{\rm NP} <
max(R_y^{-1},r_z^{-1})$.} and moderately large $\alpha$.
In the case 
of small $k$, large $\alpha$, the requirement $\epsilon <$ 0.1 causes the 
typical values $C_{np}$ to be lower than the equivalent 5-dimensional RS 
coupling. As a result the graviton production cross-section in the 
6-dimensional model would be lower than that in the 5-dimensional model, 
and, in fact, is likely to be more compatible with the observed excess. 
For the choice $\zeta$=7, we plot this favoured sector of the parameter 
space in Fig.\ref{fig:750GeVsmallk} in (a) the $m_{10} - |C_{10}|$ plane, 
and (b) the $m_{10} - k_5/\mpl$ plane. The relation between 
$k_5/\mpl$ and $C_{10}$ was noted earlier in Eq.\ref{eq:5D6D}. 
Fig.\ref{fig:750GeVsmallk}(c) shows the same region in the $k - \alpha$ plane. 
Note that the value of $\zeta$ is chosen for illustrative purposes. While it 
is indicative of the likely order of magnitude of the quantity, it is not 
a special or critical or 'best-fit' value. 

\begin{figure}[!ht]
\centering
\includegraphics[scale=0.27]{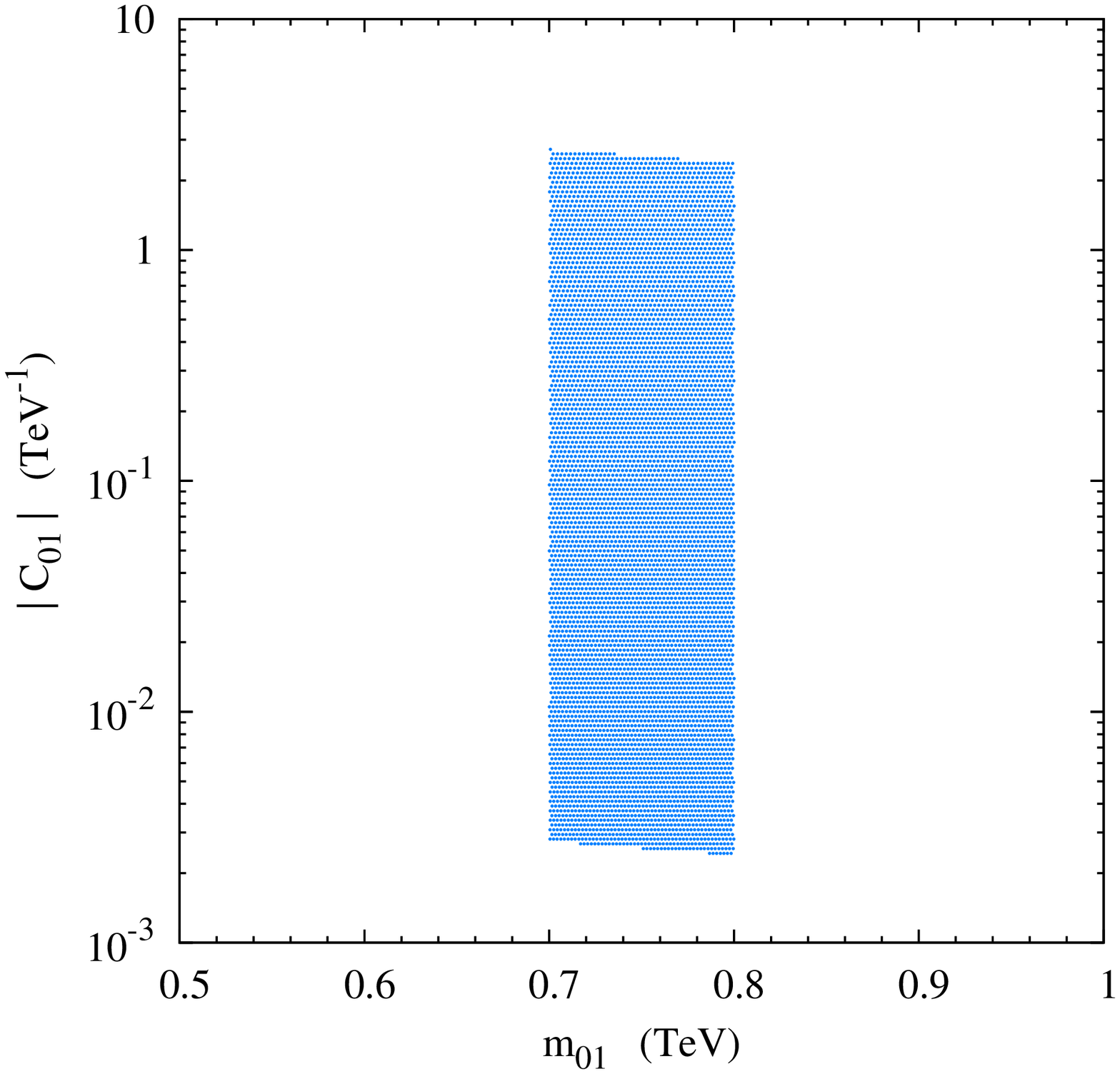}
\includegraphics[scale=0.27]{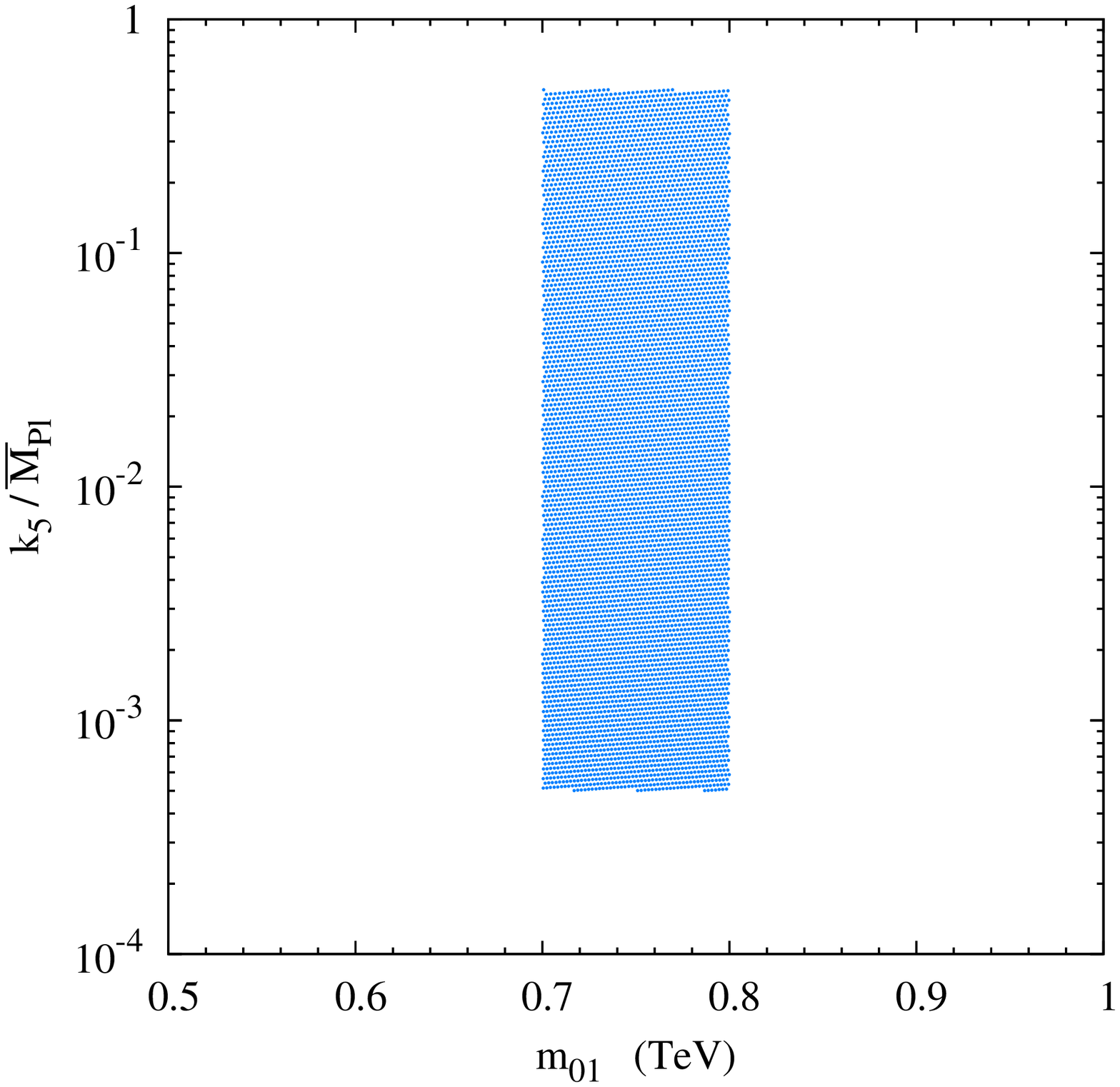}
\includegraphics[scale=0.28]{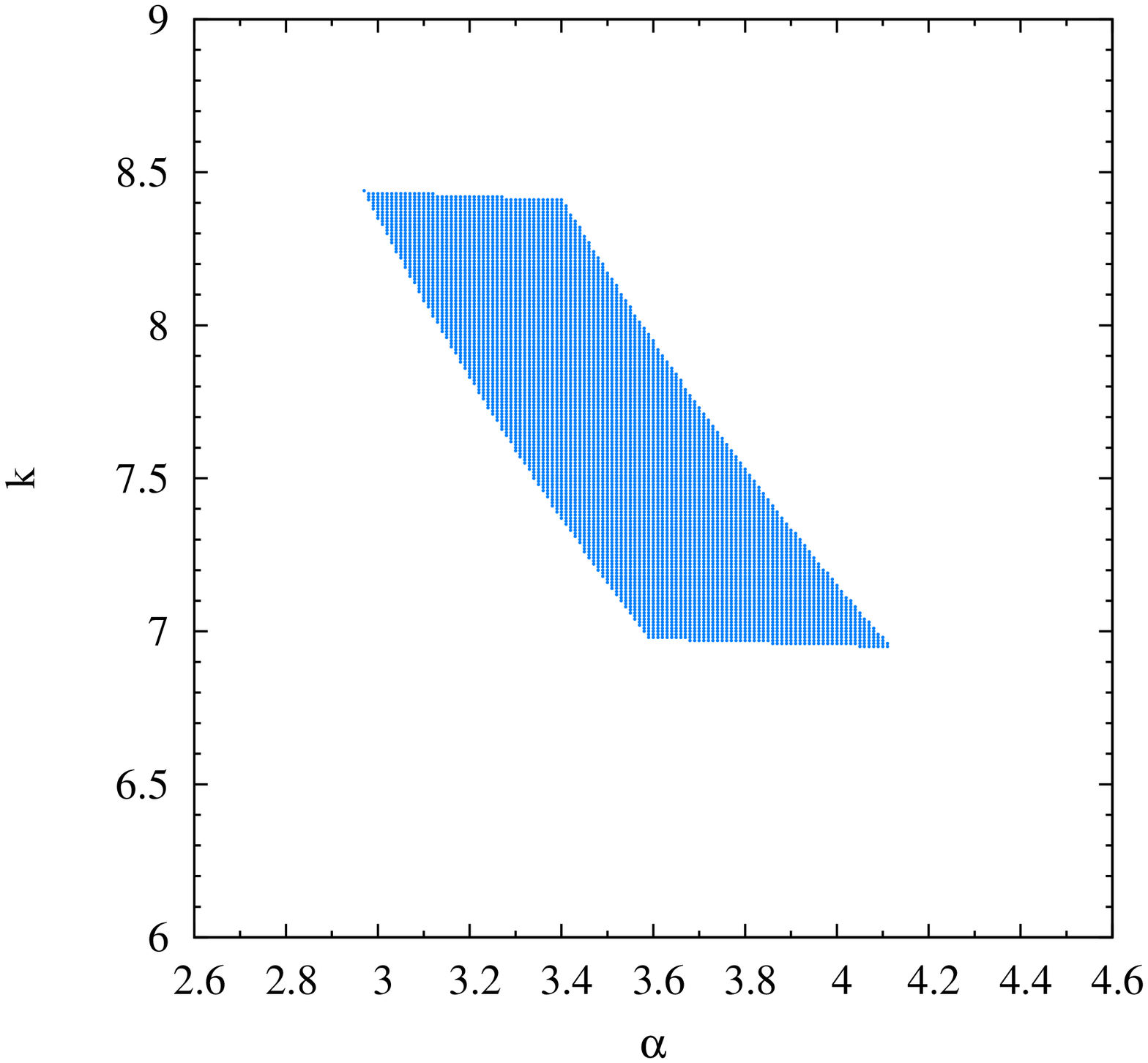}
\vskip -20pt
\caption{As in Fig.\ref{fig:750GeVsmallk}, but for 
large $k$, large $\alpha$ and $\Lambda_{\rm NP}=10 R_y^{-1}$.
}
\label{fig:750GeVlargek}
\end{figure}

Turning to the large $k$, large $\alpha$ case, we find that there exist 
sectors in the parameter space where $m_{01} \in$ [700, 800] $\gev$ and 
$C_{01}$ lies in the region close to $k_5/\mpl = 0.01$. 
In Fig.\ref{fig:750GeVlargek} we plot this sector of parameter 
space in the $m_{01} - |C_{01}|$, $m_{01} - k_5/\mpl$ and the 
$k - \alpha$ planes. This time we assume $\zeta$=10.

Clearly these regions of parameter space are neither fine-tuned nor do they 
involve large hierarchies between $R_y$ and $r_z$. In fact, they provide 
a rather satisfactory explanation for the observed deviation from the SM.
Should the observation of a resonance be confirmed with more data, 
further exploration of this region of parameter space would be in order.

%%%%%%%%%%%%%%%%%%%%%%%%%%%%%%%%%%%%%%%%%%%%%%%%%%%%%%%%%%%%%
% Figures-II
%%%%%%%%%%%%%%%%%%%%%%%%%%%%%%%%%%%%%%%%%%%%%%%%%%%%%%%%%%%%%

%%%%%%%%%%%%%%%%%%%%%%%%%%%%%%%%%%%%%%%%%%%%%%%%%%%%%%%%%%%%%
% End of Figures-II
%%%%%%%%%%%%%%%%%%%%%%%%%%%%%%%%%%%%%%%%%%%%%%%%%%%%%%%%%%%%%

\subsection{Corroborative signals from other channels}
We now return to the postponed question of the lack of signals 
in other channels. Since gravitons have a universal coupling to all brane-localized SM particles, 
one would expect that the excess in the diphoton channel would be accompanied 
by excesses corresponding to the same invariant mass in the dilepton, dijet, 
$WW$ and $ZZ$ channels. However, none such have been reported as yet.
In their updated results on the diphoton channel, the ATLAS collaboration 
reports~\cite{ATLASMar2016} an excess of 25 events in the invariant mass range 700 - 840 GeV. 
Assuming that this accurately represents the expectations from a
graviton excitation (in a theory where all the SM fields are localized on the IR-brane), this would
translate to ~ 390 additional events in the diject channel, ~12 additional events in each of the dielectron
and dimuon channel, ~42 additional events in the WW-channel and ~21 additional events in the ZZ channels
( for the last two, all decay modes of the gauge boson have been summed over). It should be 
realized though that these numbers are only indicative (derived as they are with kinematic restrictions
identical to those enforced in the diphoton channel) and would change when the actual analysis 
cuts are imposed instead. We now discuss each of them in turn.

\begin{itemize}
\item {\bf Dijet~\cite{ATLAS_dijet_13TeV,CMS_dijet_13TeV}}:      
While the analyses focus on $m_{jj} > 1.1$ $\tev$, it can be seen that for $m_{jj}\in [700,800]$ $\gev$, 
the SM expectation is in excess fo $10^5$ events. An excess of 390 events would correspond to a 
small significance ($S/\sqrt{B} < 1$).

\item {\bf $WW$ and $ZZ$~\cite{ATLAS_diboson_13TeV,CMS_diboson_13TeV}}:
The searches conducted by ATLAS are in modes where at least one of the $W$'s or $Z$'s decays into leptons, leading to a further 
suppression of the signal due to the small branching ratio of $W$ and $Z$ into leptons. 
Consequently, the lack of the signal so far is only to be expected. And while CMS does consider hadronic
decays of the W and Z, they have, yet, considered only invariant mass above 1 $\tev$. Given the fact the SM background are larger for lower invariant masses, it requires more statistics to resolve the excess in this channel. 
This situation is in marked contrast with the case of a spin-0 resonance, where, for the simplest models, decay into the diphoton channel tends to be significantly suppressed with respect to the decay into $W^+W^-$ and $ZZ$

\item {\bf Dilepton~\cite{ATLAS_dilepton_13TeV,CMS_dilepton_13TeV}}: 
In the dielectron channel, the background expectation is approximately 53 events. 
An additional 12 events coming from a graviton decay would only result in low signal 
significance with $S/\sqrt{B} \sim$ 1.6. A similar argument holds for dimuon production.  
\end{itemize}

In other words, the absence of excesses in the dijet, WW, ZZ dilepton channels 
is not yet really worrisome at least, at present.
It might be argued though that while the individual negative results are not
bothersome, in totality they present a strong counterargument to the hypothesis of a 750 $\gev$
graviton with the 6-dimensional nested warping. Indeed if additional data continues to project 
the same features as the current one, the simple model that presented here would be under threat and 
a suitable mechanism should be formulated. We turn to this now.

\subsection{An alternative scenario}

The primary problem with the graviton interpretation for an excess confined to a single channel arises from the fact that the 
branching fractions of a universally coupling graviton are uniquely determined.
Deviations from universality are possible, though, if the fermions and gauge bosons have different wave profiles in the extra-dimensions. 
This can be achieved by a minimal extension allowing SM particles to propagate into the bulk.

While such an extension into the entire bulk has been considered in Ref.~\cite{Arun:2015kva}, 
we restrict ourselves to a simpler scenario wherein the SM fields are five-dimensional entities 
rather than full six-dimensional ones.
Apart from offering the simplest extension that solves the problem at hand, the construction
presented below is a novel one.
There are two special locations
where the SM 4-brane could exist, namely as a hypersurface at $z=0$ or one at $y=\pi$.
The choice depends on the extent of warping associated with the two directions which, in turn,
play a pivotal role in defining the wavefunction and the consequent hierarchy in the fermion
masses on the one hand and the coupling to the putative 750 $\gev$ graviton on the other.
We choose $z=0$ and $y=\pi$ branes for small $k$ and large $k$ respectively. It should also be appreciated that, with the SM
fields now being five-dimensional, $\zeta >1$ is natural.

It is well known that, in the case of a five-dimensional Standard model in a Randall-Sundrum
background, the zero mode of light fermions (heavy fermions) are localized dynamically near 
the UV (IR) 3-brane, with the degree of localization controlled by the 
bulk Dirac mass term~\cite{Huber:2000ie, Grossman:1999ra, Gherghetta:2000qt}.
This serves to explain the fermion mass hierarchy.
On the other hand, the gauge boson zero modes have a flat profile in the extra-dimension. 
This difference between gauge bosons and fermions along with the fact that KK gravitons, 
(except the zero-mode graviton) are localized near the IR brane 
can engender a suppression in the graviton decay width to dileptons 
in comparison with the decay to diphotons.

As in previous sections, we are posed with two distinct regimes, namely, 
large $k$ and the other small $k$. 
Having large $k$, along with bulk SM fields, leads to large, non-perturbative
gauge boson-fermion couplings which is phenomenologically disfavoured~\cite{Arun:2015kva}. 
Hence we concentrate on the small $k$
scenario, with
the SM particles propagating on the 4-brane located at $z=0$ with the line element given by 
\[
ds_5^2 = e^{- 2 c \,  |y|}\eta_{\mu \nu}dx^{\mu}dx^{\nu} + R_y^2 dy^2 \, .
\] 

The gauge fields can be decomposed into KK-towers of 4-dimensional fields, with the $y-$dependent
factor int he wavefunction being given in terms of 
\[
\Psi_{v}^{(n)}(y) = \frac{1}{N_n} e^{c y} \Big( J_{1}(\frac{m_n R_y}{c}e^{cy }) + \beta_n Y_{1}(\frac{m_n R_y}{c}e^{cy })   \Big)
\]
In particular, the zero-mode ( to be identified with the SM field) have a simpler form
\[
\Psi_{v}^{(0)}(y) = \frac{1}{\sqrt{\pi}}
\]
and, consequently, the coupling of the graviton to a pair of vector bosons ( W, Z) remains 
unchanged\footnote{A small change does occur once the electroweak symmetry is broken 
by a brane localized Higgs field, but is of no consequence in the present context.}.

\begin{figure}[!h]
\centering
\includegraphics[scale=0.3]{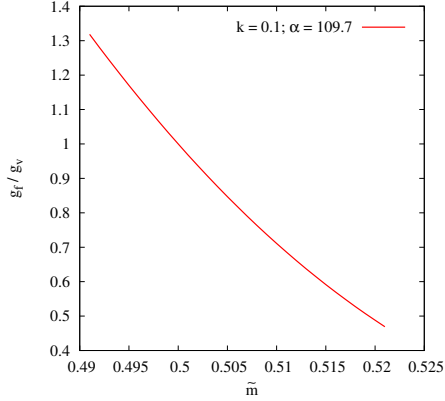}
\vskip -20pt
\caption{The ratio of the coupling of the first graviton KK-mode
to fermions ($g_f$) and gauge bosons ($g_v$).
}
\label{fig:smallkgravitoncoupling}
\end{figure}

As for the fermions, the very fact of them being vector like\footnote{Note that the 
wrong chiralities are naturally projected out by the orbifolding conditions.} allows bulk mass
terms viz, $m_D \bar{\psi_D}\psi_D + m_S \bar{\psi_S}\psi_S$ where $\psi_D$ and $\psi_S$ refers 
to $SU(2)_{L}$ doublets and singlets respectively, apart from the brane localized terms 
occurring from spontaneous symmetry breaking.
Neglecting the latter ( on account of them being much smaller than $m_D$ or $m_S$, the natural 
scale for these being $R_y^{-1}$), the wavefunction for the zero-mode can be seen to be
\[
\Psi_f(y) = \sqrt{c\frac{1-2 \tilde m}{e^{(1-2 \tilde m)c \pi}-1}} e^{(2-\tilde{m})c y}
\]
where $\tilde m = m r_c/c$ with $m = m_D, m_S$ as the case may be. Once again the calculation
of the graviton coupling is straight forward. In Fig.~\ref{fig:smallkgravitoncoupling} we
display the ratio of the graviton's coupling to fermions ($g_f$) to that with gauge bosons ($g_v$); asserting $m_D$ and $m_S$ to be equal and independent of fermion's identity. 
and it can be seen that it is not difficult to obtain a suppression large enough to evade 
the constraints from the dilepton decay channel. Indeed, with small variations in $m_D, m_S$, differing 
fermionic branching ratios can be easily accommodated were such a thing to be demanded by future measurements.

%%%%%%%%%%%%%%%%%%%%%%%%%%%%%%%%%%%%%%%%%%%%%%%%%
%%%%%%%%%%%%%%%%%%%%%%%%%%%%%%%%%%%%%%%%%%%%%%%%%

It should be noted that with the Standard Model fields being in the 5-dimensional bulk, 
the requirements of custodial symmetry and consistency with electroweak precision measurements
mean that the first KK gauge boson mass has to be greater than $3$ \tev.
This leads us into trouble, since the mass of the first KK graviton mode 
has to be greater than the first KK
gauge boson mass by a factor $1.59$, and it debars the graviton from acquiring a mass of 750 GeV. The resolution to this conundrum is to incorporate
4-brane localized Einstein-Hilbert action~\cite{Davoudiasl:2003zt,Hewett:2016omf}.
For small $k$ we choose to localize the five-dimensional 
Einstein-Hilbert term on $y=0$ and $y=\pi$ 4-branes.
The total action including the brane localized terms is
\[
S_g = M_6^4 \int d^4 x \int dy \int dz \Big(\sqrt{-g} R^{(6)} + \sqrt{-g_5} \, R_y \, \{ \,  g_0  \, \delta(y) \, + \, g_\pi \,  \delta( y - \pi)  \,  \}  R^{(5)} \Big) \, ,
\]
where $\sqrt{-g} = a^4 b^5 R_y r_z$ and $\sqrt{-g_5} = a^4 b^4 r_z$. $g_0$ and $g_\pi$ are numerical
coefficients that denote the strengths of the brane localized kinetic terms. The origin of such brane
could be quantum mechanical in nature~\cite{Georgi:2000ks}, and here we 
choose to work with the lowest order in $R$, as this will be dominant contributor. 

The relevant part of the action, since we are interested only in the 
spectrum of $h_{\mu \nu}$, could be written as
\[
S_g = M_6^4 \int d^4 x \int dy \int dz \Big(\sqrt{-g} \,  \partial^M h_{\mu \nu} \, \partial_M h_{\mu \nu} + \sqrt{-g_5} \, R_y \, \{ \, g_0 \,  \delta(y) \, + \, g_\pi \,  \delta( y - \pi)  \,  \}  \partial^{\bar M} h_{\mu \nu}\partial_{\bar M}  \, h_{\mu \nu} \Big) \,
\]
with $M=(0,1,2,3,4,5)$ and $\bar M = (0,1,2,3,5)$

\begin{figure}[!h]
\centering
\includegraphics[scale=0.3]{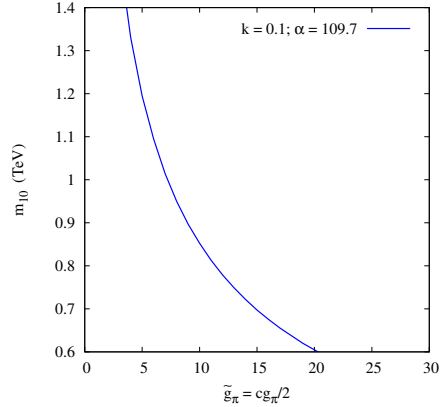}
\vskip -20pt
\caption{Variation in the mass of the first KK-mode graviton with respect to
$\frac{1}{2} c g_\pi $, for $k=0.1$ and $\alpha=109.7$ keeping $\epsilon=0.1$.
The mass of the first KK-mode of the gauge boson is held at $M_A \sim 3 \tev$.
For \mbox{$m_{10}$ = 0.75 TeV,} $\frac{1}{2} c g_0 $ = 40, $\frac{1}{2} c g_\pi $ = 13 
and $|C_{10}| = 3 \times 10^{-3} \tev^{-1}$.}
\label{grav_modified}
\end{figure}

With $g_0 $ and $g_\pi $ proportional to $b(z)$, the modified graviton masses are, as derived
in ~\cite{Davoudiasl:2003zt}, given by
\beq
\label{grav_mod}
\xi_{1}(\alpha_{10}) - \frac{1}{2} g_\pi \, c \, \alpha_{10} \, \xi_{2}(\alpha_{10}) =0 \, ,
\eeq
where $\alpha_{10} = \frac{m_{10} R_y}{c}e^{c \pi}$ and 
\[
\xi_{q} = J_{q}(\alpha_{10} \, e^{c (|y|-\pi)}) + \beta_{10} \, Y_{q}(\alpha_{10} \, e^{c (|y|-\pi)}) \, , \quad(q=1,2)
\]
with $\beta_{10} $ given as
\[
\beta_{10} = -\frac{J_{1}(\alpha_{10} e^{-c \pi}) + g_0 \frac{c R_y}{2} \alpha_{10} e^{-c \pi} J_{2}(\alpha_{10}e^{-c \pi})}{Y_{1}(\alpha_{10} e^{-c \pi}) + g_0 \frac{c R_y}{2} \alpha_{10} e^{-c \pi} Y_{2}(\alpha_{10}e^{-c \pi})}\, .
\]
Note that for $m_{10}$ of the order $\tev$, 
the constant of integration $\beta_{10} \ll 1$, and hence 
in Eq.~\ref{grav_mod} we could safely ignore
the contribution from Bessel Y function. The mass spectrum is independent of $g_0$, and with small
$g_\pi$, the spectrum tends to the root of $J_1$ as expected.
Using the relation in Eq.~\ref{grav_mod} the modified mass for the first KK mode of graviton could
be calculated for different values of $\frac{1}{2} c g_\pi $. This is plotted in
Fig.~\ref{grav_modified}, where it is easy to see that for a suitable value of $k$ we do not need
a large $g_\pi$ to achieve $750 \gev$ graviton mass. Moreover, the constraint on the mass of the first KK mode of gauge bosons from electroweak precision
data is satisfied.

%%%%%%%%%%%%%%%%%%%%%%%%%%%%%%%%%%%%%%%%%%%%%%%%%%%%%%%%%%%%%
% End of The 750 GeV excess
%%%%%%%%%%%%%%%%%%%%%%%%%%%%%%%%%%%%%%%%%%%%%%%%%%%%%%%%%%%%%

%%%%%%%%%%%%%%%%%%%%%%%%%%%%%%%%%%%%%%%%%%%%%%%%%%%%%%%%%%%%%%%%%%%%%%%
% End of Constraining the Parameter Space
%%%%%%%%%%%%%%%%%%%%%%%%%%%%%%%%%%%%%%%%%%%%%%%%%%%%%%%%%%%%%%%%%%%%%%%

%%%%%%%%%%%%%%%%%%%%%%%%%%%%%%%%%%%%%%%%%%%%%%%%%%%%%%%%%%%%%%%%%%%%%%%
% Discussion and Summary
%%%%%%%%%%%%%%%%%%%%%%%%%%%%%%%%%%%%%%%%%%%%%%%%%%%%%%%%%%%%%%%%%%%%%%%
\section{Discussion and Summary}
\label{sec:discussion}
The search for extra dimensions by ATLAS and CMS along with the discovery of 
a 125 $\gev$ Higgs boson at the LHC have diminished the parameter space of 
the 5-dimensional Randal-Sundrum model. An 
alternative minimal extension of the Randall-Sundrum model was proposed 
in Ref.~\cite{dcssg} which allowed 
for a light Higgs inspite of the gravitons being considerably heavy.
This was achieved by admitting a doubly warped 6-dimensional manifold with four 
4-branes protecting the edges at the orbifold fixed points. The existence of 
fifth spatial dimension introduces some extra parameters (though 
not all independent) in the form of modulus hierarchy and warp factors. 
In this paper we have identified the regions in the parameter space of this 
model that survive the current LHC constraints and those that can tested during Run 2 of LHC. 

We have examined three regimes, namely large $k$ large $\alpha$, 
large $k$ small $\alpha$ and small $k$ large $\alpha$. The latter is 
unconstrained by present LHC data since the KK gravitons in the parameter 
space are heavier than the energy scale probed by 8 $\tev$ LHC. But the 
large $k$ scenario gets constrained and this is shown in 
Fig.\ref{fig:ATLAS_largeK_largeA} and Fig.\ref{fig:ATLAS_largeK_smallA}

$k >$ 8 is disfavoured as they lead the graviton's width
to be larger than its mass whereas $k <$ 0.8 is inaccessible at the LHC.
Other than this, comparison with predictions for the Drell-Yan and diphoton processes at 14 $\tev$ with 
50 fb$^{-1}$ worth of data~\cite{Das:2014tva} seem to imply that a
wide range of $\alpha$ values can be accommodated across the two 
$k$ regimes. 

Finally we have delineated the region of parameter space that can explain the 
reported excess in the $m_{\gamma\gamma}$ distribution measured in the 13 $\tev$ 
run of the LHC. The significance of the excess is, at present, rather small. 
However, in case this significance increases with accumulation of more data, 
the 6-dimensional multiply warped model discussed here would certainly 
make for a compelling explanation for, on the one hand the model has several 
interesting features, and, on the other, a 750 GeV graviton comes about 
quite naturally without stretching the parameter space.
We have also outlined a possible mechanism that would 
naturally give rise to a low mass graviton with dilepton couplings
being suppressed
in comparison to diphoton couplings. In case, even after the
accumulation of more data, no excess is seen in the dilepton channel, this scenario will
assume greater importance. 

%%%%%%%%%%%%%%%%%%%%%%%%%%%%%%%%%%%%%%%%%%%%%%%%%%%%%%%%%%%%%%%%%%%%%%%
% End of Discussion and Summary
%%%%%%%%%%%%%%%%%%%%%%%%%%%%%%%%%%%%%%%%%%%%%%%%%%%%%%%%%%%%%%%%%%%%%%%

%%%%%%%%%%%%%%%%%%%%%%%%%%%%%%%%%%%%%%%%%%%%%%%%%%%%%%%%%%%%%%%%%%%%%%%
% Acknowledgements
%%%%%%%%%%%%%%%%%%%%%%%%%%%%%%%%%%%%%%%%%%%%%%%%%%%%%%%%%%%%%%%%%%%%%%%
\section*{Acknowledgement}
MTA would like to thank UGC-CSIR, India for assistance under 
Senior Research Fellowship Grant Sch/SRF/AA/139/F-123/2011-12.
PS acknowledges support from the Department of Science and Technology, 
India and thanks the IRC, University of Delhi 
and RECAPP, Harish-Chandra Research Institute for hospitality and 
computational facilities during different stages of this work.
%%%%%%%%%%%%%%%%%%%%%%%%%%%%%%%%%%%%%%%%%%%%%%%%%%%%%%%%%%%%%%%%%%%%%%%
% End of Acknowledgements
%%%%%%%%%%%%%%%%%%%%%%%%%%%%%%%%%%%%%%%%%%%%%%%%%%%%%%%%%%%%%%%%%%%%%%%

%%%%%%%%%%%%%%%%%%%%%%%%%%%%%%%%%%%%%%%%%%%%%%%%%%%%%%%%%%%%%%%%%%%%%%%
% References
%%%%%%%%%%%%%%%%%%%%%%%%%%%%%%%%%%%%%%%%%%%%%%%%%%%%%%%%%%%%%%%%%%%%%%%

%%%%%%%%%%%%%%%%%%%%%%%%%%%%%%%%%%%%%%%%%%%%%%%%%%%%%%%%%%%%%%%%%%%%%%%
% End of References
%%%%%%%%%%%%%%%%%%%%%%%%%%%%%%%%%%%%%%%%%%%%%%%%%%%%%%%%%%%%%%%%%%%%%%%


\begin{thebibliography}{99}

\bibitem{RS}
%\cite{Randall:1999ee}
%\bibitem{Randall:1999ee} 
  L.~Randall and R.~Sundrum,
  %``A Large mass hierarchy from a small extra dimension,''
  Phys.\ Rev.\ Lett.\  {\bf 83}, 3370 (1999);
%  [hep-ph/9905221];\\
  %%CITATION = HEP-PH/9905221;%%
{\it ibid} {\bf 83}, 4690 (1999). 
% [hep-th/9906064].

\bibitem{GW1} 
%\cite{Goldberger:1999uk} 
%\bibitem{Goldberger:1999uk} 
  W.D.~Goldberger and M.~B.~Wise,
  %``Modulus stabilization with bulk fields,''
  Phys.\ Rev.\ Lett.\  {\bf 83}, 4922 (1999).
%  [hep-ph/9907447].
  %%CITATION = HEP-PH/9907447;%%

\bibitem{Green}
M.B.~Green, J.H.~Schwarz and E.~Witten, ``{\em Superstring Theory}'',Vols.I \&
II, Cambridge University Press (1987);
J.~Polchinski, ``{\em String Theory}'', Cambridge University Press (1998).

\bibitem{diboson}
%\cite{Aad:2012cy}
%\bibitem{Aad:2012cy}
  G.~Aad {\it et al.} [ATLAS Collaboration],
  %``Search for Extra Dimensions in diphoton events using proton-proton collisions recorded at $\sqrt{s}=7$ TeV with the ATLAS detector at the LHC,''
  New J.\ Phys.\  {\bf 15} (2013) 043007;
  %doi:10.1088/1367-2630/15/4/043007
%  [arXiv:1210.8389 [hep-ex]];\\
%\cite{Aad:2015owa}
%\bibitem{Aad:2015owa}
%  G.~Aad {\it et al.} [ATLAS Collaboration],
  %``Search for high-mass diboson resonances with boson-tagged jets in proton-proton collisions at $\sqrt{s} = 8$ TeV with the ATLAS detector,''
  arXiv:1506.00962 [hep-ex].
  
%\cite{CMS:2015cwa}
%\bibitem{CMS:2015cwa}
\bibitem{EXO-12-045}
  The CMS Collaboration, % [CMS Collaboration],
  %``Search for High-Mass Diphoton Resonances in pp Collisions at sqrt(s)=8 TeV with the CMS Detector,''
  CMS-PAS-EXO-12-045.

\bibitem{dilepton}
%\cite{Aad:2014cka}
%\bibitem{Aad:2014cka}
  G.~Aad {\it et al.} [ATLAS Collaboration],
  %``Search for high-mass dilepton resonances in pp collisions at $\sqrt{s}=8$  TeV with the ATLAS detector,''
  Phys.\ Rev.\ D {\bf 90} (2014) 5,  052005;
  %doi:10.1103/PhysRevD.90.052005
%  [arXiv:1405.4123 [hep-ex]];\\
%\cite{Khachatryan:2014fba}
%\bibitem{Khachatryan:2014fba}
  V.~Khachatryan {\it et al.} [CMS Collaboration],
  %``Search for physics beyond the standard model in dilepton mass spectra in proton-proton collisions at $ \sqrt{s}=8 $ TeV,''
  JHEP {\bf 1504} (2015) 025;
  %doi:10.1007/JHEP04(2015)025
%  [arXiv:1412.6302 [hep-ex]];\\
%\cite{Khachatryan:2014gha}
%\bibitem{Khachatryan:2014gha}
%  V.~Khachatryan {\it et al.} [CMS Collaboration],
  %``Search for massive resonances decaying into pairs of boosted bosons in semi-leptonic final states at $\sqrt{s} =$ 8 TeV,''
  JHEP {\bf 1408} (2014) 174.
  %doi:10.1007/JHEP08(2014)174
%  [arXiv:1405.3447 [hep-ex]].

%\cite{ATLAS:CONF-2015-081}
%\bibitem{ATLAS:CONF-2015-081}
\bibitem{ATLASDec2015}
  The ATLAS Collaboration, % [ATLAS Collaboration],
  %``Search for resonances decaying to photon pairs in 3.2 fb$^{-1}$ of $pp$ collisions at $\sqrt{s}$ = 13 TeV with the ATLAS detector,"
      ATLAS-CONF-2015-081.

%\cite{CMS:2015dxe}
%\bibitem{CMS:2015dxe}
\bibitem{CMSDec2015}
  The CMS Collaboration, % [CMS Collaboration],
  %``Search for new physics in high mass diphoton events in proton-proton collisions at 13TeV,''
  CMS-PAS-EXO-15-004.
  %%CITATION = CMS-PAS-EXO-15-004;%%
  %8 citations counted in INSPIRE as of 17 Dec 201

\bibitem{ATLAS2015}  
%\cite{Aad:2015mna}
%\bibitem{Aad:2015mna} 
  G.~Aad {\it et al.} [ATLAS Collaboration],
  %``Search for high-mass diphoton resonances in $pp$ collisions at $\sqrt{s}=8$ TeV with the ATLAS detector,''
  Phys.\ Rev.\ D {\bf 92}, 032004 (2015).
%  [arXiv:1504.05511 [hep-ex]].
  %%CITATION = ARXIV:1504.05511;%%
  %6 citations counted in INSPIRE as of 03 Nov 2015
  
\bibitem{RSextn_previous}
%\cite{Gogberashvili:2003xa}
%\bibitem{Gogberashvili:2003xa}
  M.~Gogberashvili and D.~Singleton,
  %`{\em `Nonsingular increasing gravitational potential for the brane in 6-D}''
  Phys.\ Lett.\ B {\bf 582}, 95 (2004); 
%  [hep-th/0310048];\\
  %%CITATION = HEP-TH/0310048;%%
  %36 citations counted in INSPIRE as of 10 Feb 2015
%
%\cite{Gogberashvili:2003ys}
%\bibitem{Gogberashvili:2003ys}
%  M.~Gogberashvili and D.~Singleton,
  %``{\em Brane in 6-D with increasing gravitational trapping potential}''
  Phys.\ Rev.\ D {\bf 69}, 026004 (2004);
%  [hep-th/0305241];\\
  %%CITATION = HEP-TH/0305241;%%
  %48 citations counted in INSPIRE as of 10 Feb 2015
%  
%\cite{Gogberashvili:2007gg}
%\bibitem{Gogberashvili:2007gg}
  M.~Gogberashvili, P.~Midodashvili and D.~Singleton,
  %``{\em Fermion Generations from 'Apple-Shaped' Extra Dimensions}''
  JHEP {\bf 0708}, 033 (2007).
%  [arXiv:0706.0676 [hep-th]].
  %%CITATION = ARXIV:0706.0676;%%
  %41 citations counted in INSPIRE as of 10 Feb 2015
 
\bibitem{6D-Salvio} 
%\cite{Parameswaran:2006db}
%\bibitem{Parameswaran:2006db} 
  S.~L.~Parameswaran, S.~Randjbar-Daemi and A.~Salvio,
  %``Gauge Fields, Fermions and Mass Gaps in 6D Brane Worlds,''
  Nucl.\ Phys.\ B {\bf 767}, 54 (2007);
%  doi:10.1016/j.nuclphysb.2006.12.020
%  [hep-th/0608074];\\
  %%CITATION = doi:10.1016/j.nuclphysb.2006.12.020;%%
  %75 citations counted in INSPIRE as of 07 Jan 2016
%
%\cite{Salvio:2009mp}
%\bibitem{Salvio:2009mp} 
  A.~Salvio,
  %``Brane Gravitational Interactions from 6D Supergravity,''
  Phys.\ Lett.\ B {\bf 681}, 166 (2009).
%  doi:10.1016/j.physletb.2009.10.008
%  [arXiv:0909.0023 [hep-th]].
  %%CITATION = doi:10.1016/j.physletb.2009.10.008;%%
  %9 citations counted in INSPIRE as of 07 Jan 2016
 
\bibitem{dcssg}
%\cite{Choudhury:2006nj}
%\bibitem{Choudhury:2006nj} 
  D.~Choudhury and S.~SenGupta,
  %``Living on the edge in a spacetime with multiple warping,''
  Phys.\ Rev.\ D {\bf 76}, 064030 (2007).
%  [hep-th/0612246].
  %%CITATION = HEP-TH/0612246;%%

%\cite{Arun:2014dga}
\bibitem{Arun:2014dga} 
  M.~T.~Arun, D.~Choudhury, A.~Das and S.~SenGupta,
  %M.T.~Arun {\it et al.},
  %``Graviton modes in multiply warped geometry,''
  Phys.\ Lett.\ B {\bf 746}, 266 (2015).
%  [arXiv:1410.5591 [hep-ph]].

%\cite{Das:2014tva}
\bibitem{Das:2014tva} 
  G.~Das, P.~Mathews, V.~Ravindran and S.~Seth,
  %G.~Das {\it et al.},
  %``RS resonance in di-final state production at the LHC to NLO+PS accuracy,''
  JHEP {\bf 1410}, 188 (2014). 
%  [arXiv:1408.3970 [hep-ph]].

\bibitem{diphoton_deluge}
%\cite{Chao:2015nsm}
%\bibitem{Chao:2015nsm} 
  W.~Chao,
  %``Symmetries Behind the 750 GeV Diphoton Excess,''
  arXiv:1512.06297 [hep-ph];
  %%CITATION = ARXIV:1512.06297;%%
  %47 citations counted in INSPIRE as of 07 Jan 2016
%
%\cite{Bernon:2015abk}
%\bibitem{Bernon:2015abk} 
  J.~Bernon and C.~Smith,
  %``Could the width of the diphoton anomaly signal a three-body decay ?,''
  arXiv:1512.06113 [hep-ph];
  %%CITATION = ARXIV:1512.06113;%%
  %48 citations counted in INSPIRE as of 07 Jan 2016
%
%\cite{Carpenter:2015ucu}
%\bibitem{Carpenter:2015ucu} 
  L.~M.~Carpenter {\it et al.}  %, R.~Colburn and J.~Goodman,
  %``Supersoft SUSY Models and the 750 GeV Diphoton Excess, Beyond Effective Operators,''
  arXiv:1512.06107 [hep-ph];
  %%CITATION = ARXIV:1512.06107;%%
  %52 citations counted in INSPIRE as of 07 Jan 2016
%
%\cite{Megias:2015ory}
%\bibitem{Megias:2015ory} 
  E.~Megias {\it et al.}  %, O.~Pujolas and M.~Quiros,
  %``On dilatons and the LHC diphoton excess,''
  arXiv:1512.06106 [hep-ph];
  %%CITATION = ARXIV:1512.06106;%%
  %52 citations counted in INSPIRE as of 07 Jan 2016
%
%\cite{Alves:2015jgx}
%\bibitem{Alves:2015jgx} 
  A.~Alves {\it et al.}  %, A.~G.~Dias and K.~Sinha,
  %``The 750 GeV $S$-cion: Where else should we look for it?,''
  arXiv:1512.06091 [hep-ph];
  %%CITATION = ARXIV:1512.06091;%%
  %59 citations counted in INSPIRE as of 07 Jan 2016
%
%\cite{Kim:2015ron}
%\bibitem{Kim:2015ron} 
  J.~S.~Kim {\it et al.}  %, J.~Reuter, K.~Rolbiecki and R.~R.~de Austri,
  %``A resonance without resonance: scrutinizing the diphoton excess at 750 GeV,''
  arXiv:1512.06083 [hep-ph];
  %%CITATION = ARXIV:1512.06083;%%
  %50 citations counted in INSPIRE as of 07 Jan 2016
%
%\cite{Ghosh:2015apa}
%\bibitem{Ghosh:2015apa} 
  S.~Ghosh {\it et al.}  %, A.~Kundu and S.~Ray,
  %``On the potential of a singlet scalar enhanced Standard Model,''
  arXiv:1512.05786 [hep-ph];
  %%CITATION = ARXIV:1512.05786;%%
  %5 citations counted in INSPIRE as of 07 Jan 2016
%
%\cite{Bai:2015nbs}
%\bibitem{Bai:2015nbs} 
  Y.~Bai {\it et al.}  %, J.~Berger and R.~Lu,
  %``A 750 GeV Dark Pion: Cousin of a Dark G-parity-odd WIMP,''
  arXiv:1512.05779 [hep-ph];
  %%CITATION = ARXIV:1512.05779;%%
  %75 citations counted in INSPIRE as of 07 Jan 2016
%
%\cite{Falkowski:2015swt}
%\bibitem{Falkowski:2015swt} 
  A.~Falkowski {\it et al.}  %, O.~Slone and T.~Volansky,
  %``Phenomenology of a 750 GeV Singlet,''
  arXiv:1512.05777 [hep-ph];
  %%CITATION = ARXIV:1512.05777;%%
  %90 citations counted in INSPIRE as of 07 Jan 2016
%
%\cite{Csaki:2015vek}
%\bibitem{Csaki:2015vek} 
  C.~Csaki {\it et al.}  %, J.~Hubisz and J.~Terning,
  %``The Minimal Model of a Diphoton Resonance: Production without Gluon Couplings,''
  arXiv:1512.05776 [hep-ph];
  %%CITATION = ARXIV:1512.05776;%%
  %82 citations counted in INSPIRE as of 07 Jan 2016
%
%\cite{Agrawal:2015dbf}
%\bibitem{Agrawal:2015dbf} 
  P.~Agrawal {\it et al.}  %, J.~Fan, B.~Heidenreich, M.~Reece and M.~Strassler,
  %``Experimental Considerations Motivated by the Diphoton Excess at the LHC,''
  arXiv:1512.05775 [hep-ph];
  %%CITATION = ARXIV:1512.05775;%%
  %66 citations counted in INSPIRE as of 07 Jan 2016
%
%\cite{Ahmed:2015uqt}
%\bibitem{Ahmed:2015uqt} 
  A.~Ahmed {\it et al.}  %, B.~M.~Dillon, B.~Grzadkowski, J.~F.~Gunion and Y.~Jiang,
  %``Higgs-radion interpretation of 750 GeV di-photon excess at the LHC,''
  arXiv:1512.05771 [hep-ph];
  %%CITATION = ARXIV:1512.05771;%%
  %74 citations counted in INSPIRE as of 07 Jan 2016
%
%\cite{Chakrabortty:2015hff}
%\bibitem{Chakrabortty:2015hff} 
  J.~Chakrabortty {\it et al.}  %, A.~Choudhury, P.~Ghosh, S.~Mondal and T.~Srivastava,
  %``Di-photon resonance around 750 GeV: shedding light on the theory underneath,''
  arXiv:1512.05767 [hep-ph];
  %%CITATION = ARXIV:1512.05767;%%
  %81 citations counted in INSPIRE as of 07 Jan 2016
%
%\cite{Bian:2015kjt}
%\bibitem{Bian:2015kjt} 
  L.~Bian {\it et al.}  %, N.~Chen, D.~Liu and J.~Shu,
  %``A hidden confining world on the 750 GeV diphoton excess,''
  arXiv:1512.05759 [hep-ph];
  %%CITATION = ARXIV:1512.05759;%%
  %82 citations counted in INSPIRE as of 07 Jan 2016
%
%\cite{Curtin:2015jcv}
%\bibitem{Curtin:2015jcv} 
  D.~Curtin and C.~B.~Verhaaren,
  %``Quirky Explanations for the Diphoton Excess,''
  arXiv:1512.05753 [hep-ph];
  %%CITATION = ARXIV:1512.05753;%%
  %75 citations counted in INSPIRE as of 07 Jan 2016
%
%\cite{Fichet:2015vvy}
%\bibitem{Fichet:2015vvy} 
  S.~Fichet {\it et al.}  %, G.~von Gersdorff and C.~Royon,
  %``Scattering Light by Light at 750 GeV at the LHC,''
  arXiv:1512.05751 [hep-ph];
  %%CITATION = ARXIV:1512.05751;%%
  %82 citations counted in INSPIRE as of 07 Jan 2016
%
%\cite{Chao:2015ttq}
%\bibitem{Chao:2015ttq} 
  W.~Chao {\it et al.}  %, R.~Huo and J.~H.~Yu,
  %``The Minimal Scalar-Stealth Top Interpretation of the Diphoton Excess,''
  arXiv:1512.05738 [hep-ph];
  %%CITATION = ARXIV:1512.05738;%%
  %78 citations counted in INSPIRE as of 07 Jan 2016
%
%\cite{Demidov:2015zqn}
%\bibitem{Demidov:2015zqn} 
  S.~V.~Demidov and D.~S.~Gorbunov,
  %``On sgoldstino interpretation of the diphoton excess,''
  arXiv:1512.05723 [hep-ph];
  %%CITATION = ARXIV:1512.05723;%%
  %76 citations counted in INSPIRE as of 07 Jan 2016
%
%\cite{No:2015bsn}
%\bibitem{No:2015bsn} 
  J.~M.~No {\it et al.}  %, V.~Sanz and J.~Setford,
  %``See-Saw Composite Higgses at the LHC: Linking Naturalness to the $750$ GeV Di-Photon Resonance,''
  arXiv:1512.05700 [hep-ph];
  %%CITATION = ARXIV:1512.05700;%%
  %80 citations counted in INSPIRE as of 07 Jan 2016
%
%\cite{Becirevic:2015fmu}
%\bibitem{Becirevic:2015fmu} 
  D.~Becirevic {\it et al.}  %, E.~Bertuzzo, O.~Sumensari and R.~Z.~Funchal,
  %``Can the new resonance at LHC be a CP-Odd Higgs boson?,''
  arXiv:1512.05623 [hep-ph];
  %%CITATION = ARXIV:1512.05623;%%
  %75 citations counted in INSPIRE as of 07 Jan 2016
%
%\cite{Cox:2015ckc}
%\bibitem{Cox:2015ckc} 
  P.~Cox {\it et al.}  %, A.~D.~Medina, T.~S.~Ray and A.~Spray,
  %``Diphoton Excess at 750 GeV from a Radion in the Bulk-Higgs Scenario,''
  arXiv:1512.05618 [hep-ph];
  %%CITATION = ARXIV:1512.05618;%%
  %79 citations counted in INSPIRE as of 07 Jan 2016
%
%\cite{Kobakhidze:2015ldh}
%\bibitem{Kobakhidze:2015ldh} 
  A.~Kobakhidze {\it et al.}  %, F.~Wang, L.~Wu, J.~M.~Yang and M.~Zhang,
  %``LHC 750 GeV diphoton resonance explained as a heavy scalar in top-seesaw model,''
  arXiv:1512.05585 [hep-ph];
  %%CITATION = ARXIV:1512.05585;%%
  %81 citations counted in INSPIRE as of 07 Jan 2016
%
%\cite{Matsuzaki:2015che}
%\bibitem{Matsuzaki:2015che} 
  S.~Matsuzaki and K.~Yamawaki,
  %``750 GeV Diphoton Signal from One-Family Walking Technipion,''
  arXiv:1512.05564 [hep-ph];
  %%CITATION = ARXIV:1512.05564;%%
  %70 citations counted in INSPIRE as of 07 Jan 2016
%
%\cite{Cao:2015pto}
%\bibitem{Cao:2015pto} 
  Q.~H.~Cao {\it et al.}  %, Y.~Liu, K.~P.~Xie, B.~Yan and D.~M.~Zhang,
  %``A Boost Test of Anomalous Diphoton Resonance at the LHC,''
  arXiv:1512.05542 [hep-ph];
  %%CITATION = ARXIV:1512.05542;%%
  %73 citations counted in INSPIRE as of 07 Jan 2016
%
%\cite{Dutta:2015wqh}
%\bibitem{Dutta:2015wqh} 
  B.~Dutta {\it et al.}  %, Y.~Gao, T.~Ghosh, I.~Gogoladze and T.~Li,
  %``Interpretation of the diphoton excess at CMS and ATLAS,''
  arXiv:1512.05439 [hep-ph];
  %%CITATION = ARXIV:1512.05439;%%
  %78 citations counted in INSPIRE as of 07 Jan 2016
%
%\cite{Molinaro:2015cwg}
%\bibitem{Molinaro:2015cwg} 
  E.~Molinaro {\it et al.}  %, F.~Sannino and N.~Vignaroli,
  %``Minimal Composite Dynamics versus Axion Origin of the Diphoton excess,''
  arXiv:1512.05334 [hep-ph];
  %%CITATION = ARXIV:1512.05334;%%
  %88 citations counted in INSPIRE as of 07 Jan 2016
%
%\cite{Petersson:2015mkr}
%\bibitem{Petersson:2015mkr} 
  C.~Petersson and R.~Torre,
  %``The 750 GeV diphoton excess from the goldstino superpartner,''
  arXiv:1512.05333 [hep-ph];
  %%CITATION = ARXIV:1512.05333;%%
  %94 citations counted in INSPIRE as of 07 Jan 2016
%
%\cite{Gupta:2015zzs}
%\bibitem{Gupta:2015zzs} 
  R.~S.~Gupta {\it et al.}  %, S.~Jäger, Y.~Kats, G.~Perez and E.~Stamou,
  %``Interpreting a 750 GeV Diphoton Resonance,''
  arXiv:1512.05332 [hep-ph];
  %%CITATION = ARXIV:1512.05332;%%
  %96 citations counted in INSPIRE as of 07 Jan 2016
%
%\cite{Bellazzini:2015nxw}
%\bibitem{Bellazzini:2015nxw} 
  B.~Bellazzini {\it et al.}  %, R.~Franceschini, F.~Sala and J.~Serra,
  %``Goldstones in Diphotons,''
  arXiv:1512.05330 [hep-ph];
  %%CITATION = ARXIV:1512.05330;%%
  %89 citations counted in INSPIRE as of 07 Jan 2016
%
%\cite{Low:2015qep}
%\bibitem{Low:2015qep} 
  M.~Low {\it et al.}  %, A.~Tesi and L.~T.~Wang,
  %``A pseudoscalar decaying to photon pairs in the early LHC run 2 data,''
  arXiv:1512.05328 [hep-ph];
  %%CITATION = ARXIV:1512.05328;%%
  %97 citations counted in INSPIRE as of 07 Jan 2016
%
%\cite{Ellis:2015oso}
%\bibitem{Ellis:2015oso} 
  J.~Ellis {\it et al.}  %, S.~A.~R.~Ellis, J.~Quevillon, V.~Sanz and T.~You,
  %``On the Interpretation of a Possible $\sim 750$ GeV Particle Decaying into $\gamma \gamma$,''
  arXiv:1512.05327 [hep-ph];
  %%CITATION = ARXIV:1512.05327;%%
  %99 citations counted in INSPIRE as of 07 Jan 2016
%
%\cite{McDermott:2015sck}
%\bibitem{McDermott:2015sck} 
  S.~D.~McDermott {\it et al.}  %, P.~Meade and H.~Ramani,
  %``Singlet Scalar Resonances and the Diphoton Excess,''
  arXiv:1512.05326 [hep-ph];
  %%CITATION = ARXIV:1512.05326;%%
  %102 citations counted in INSPIRE as of 07 Jan 2016
%
%\cite{DiChiara:2015vdm}
%\bibitem{DiChiara:2015vdm} 
  S.~Di Chiara {\it et al.} %, L.~Marzola and M.~Raidal,
  %``First interpretation of the 750 GeV di-photon resonance at the LHC,''
  arXiv:1512.04939 [hep-ph];
  %%CITATION = ARXIV:1512.04939;%%
  %102 citations counted in INSPIRE as of 07 Jan 2016
%
%\cite{Franceschini:2015kwy}
%\bibitem{Franceschini:2015kwy} 
  R.~Franceschini {\it et al.},
  %``What is the gamma gamma resonance at 750 GeV?,''
  arXiv:1512.04933 [hep-ph];
  %%CITATION = ARXIV:1512.04933;%%
  %122 citations counted in INSPIRE as of 07 Jan 2016
%
%\cite{Buttazzo:2015txu}
%\bibitem{Buttazzo:2015txu} 
  D.~Buttazzo {\it et al.} %, A.~Greljo and D.~Marzocca,
  %``Knocking on New Physics' door with a Scalar Resonance,''
  arXiv:1512.04929 [hep-ph];
  %%CITATION = ARXIV:1512.04929;%%
  %105 citations counted in INSPIRE as of 07 Jan 2016
%
%\cite{Knapen:2015dap}
%\bibitem{Knapen:2015dap} 
  S.~Knapen {\it et al.} %, T.~Melia, M.~Papucci and K.~Zurek,
  %``Rays of light from the LHC,''
  arXiv:1512.04928 [hep-ph];
  %%CITATION = ARXIV:1512.04928;%%
  %99 citations counted in INSPIRE as of 07 Jan 2016
%
%\cite{Nakai:2015ptz}
%\bibitem{Nakai:2015ptz} 
  Y.~Nakai {\it et al.} %, R.~Sato and K.~Tobioka,
  %``Footprints of New Strong Dynamics via Anomaly,''
  arXiv:1512.04924 [hep-ph];
  %%CITATION = ARXIV:1512.04924;%%
  %86 citations counted in INSPIRE as of 07 Jan 2016
%
%\cite{Backovic:2015fnp}
%\bibitem{Backovic:2015fnp} 
  M.~Backovic {\it et al.} %, A.~Mariotti and D.~Redigolo,
  %``Di-photon excess illuminates Dark Matter,''
  arXiv:1512.04917 [hep-ph];
  %%CITATION = ARXIV:1512.04917;%%
  %99 citations counted in INSPIRE as of 07 Jan 2016
%
%\cite{Harigaya:2015ezk}
%\bibitem{Harigaya:2015ezk} 
  K.~Harigaya and Y.~Nomura,
  %``Composite Models for the 750 GeV Diphoton Excess,''
  arXiv:1512.04850 [hep-ph].
  %%CITATION = ARXIV:1512.04850;%%
  %102 citations counted in INSPIRE as of 07 Jan 2016

\bibitem{SteveMyers}
  Steve Myers, ICHEP 2010, 
  https://indico.cern.ch/event/73513/session/13/contribution/73.

  \bibitem{ATLASMar2016}
%\cite{}
%\bibitem{} 
  The ATLAS collaboration,
  %``Search for resonances in diphoton events with the ATLAS detector at $\sqrt{s}$ = 13 TeV,''
  ATLAS-CONF-2016-018.
  %%CITATION = ATLAS-CONF-2016-018;%%
  %11 citations counted in INSPIRE as of 27 Apr 2016

\bibitem{ATLAS_dijet_13TeV} 
%\cite{ATLAS:2015nsi}
%\bibitem{ATLAS:2015nsi} 
  G.~Aad {\it et al.} [ATLAS Collaboration],
  %``Search for new phenomena in dijet mass and angular distributions from $pp$ collisions at $\sqrt{s}=$ 13 TeV with the ATLAS detector,''
  Phys.\ Lett.\ B {\bf 754}, 302 (2016).
  %doi:10.1016/j.physletb.2016.01.032
%  [arXiv:1512.01530 [hep-ex]].
  %%CITATION = doi:10.1016/j.physletb.2016.01.032;%%
  %43 citations counted in INSPIRE as of 27 Apr 2016
 
\bibitem{CMS_dijet_13TeV} 
%\cite{Khachatryan:2015dcf}
%\bibitem{Khachatryan:2015dcf} 
  V.~Khachatryan {\it et al.} [CMS Collaboration],
  %``Search for narrow resonances decaying to dijets in proton-proton collisions at $\sqrt(s) =$ 13 TeV,''
  Phys.\ Rev.\ Lett.\  {\bf 116}, 071801 (2016).
%  doi:10.1103/PhysRevLett.116.071801
%  [arXiv:1512.01224 [hep-ex]].
  %%CITATION = doi:10.1103/PhysRevLett.116.071801;%%
  %51 citations counted in INSPIRE as of 27 Apr 2016
  
\bibitem{ATLAS_diboson_13TeV} 
%\cite{}
%\bibitem{} 
  The ATLAS Collaboration,
  %``Search for diboson resonances in the $\nu\nu qq$ final state in $pp$ collisions at $\sqrt{s}=$13 TeV with the ATLAS detector,''
  ATLAS-CONF-2015-068;
  %%CITATION = ATLAS-CONF-2015-068;%%
  %9 citations counted in INSPIRE as of 27 Apr 2016
%\cite{}
%\bibitem{} 
%  The ATLAS collaboration,
  %``Search for diboson resonances in the llqq final state in pp collisions at $\sqrt{s}$ = 13 TeV with the ATLAS detector,''
  ATLAS-CONF-2015-071;
  %%CITATION = ATLAS-CONF-2015-071;%%
  %15 citations counted in INSPIRE as of 27 Apr 2016
%\cite{}
%\bibitem{} 
%  The ATLAS collaboration,
  %``Search for $WW/WZ$ resonance production in the $\ell\nu qq$ final state at $\sqrt{s}=13\,$ TeV with the ATLAS detector at the LHC,''
  ATLAS-CONF-2015-075.
  %%CITATION = ATLAS-CONF-2015-075;%%
  %17 citations counted in INSPIRE as of 27 Apr 2016

\bibitem{CMS_diboson_13TeV}
%\cite{CMS:2015nmz}
%\bibitem{CMS:2015nmz} 
  The CMS Collaboration, % [CMS Collaboration],
  %``Search for massive resonances decaying into pairs of boosted W and Z bosons at $\sqrt{s}$ = 13 TeV,''
  CMS-PAS-EXO-15-002.
  %%CITATION = CMS-PAS-EXO-15-002;%%
  %18 citations counted in INSPIRE as of 27 Apr 2016

\bibitem{ATLAS_dilepton_13TeV}  
%\cite{}
%\bibitem{} 
  The ATLAS collaboration,
  %``Search for new phenomena in the dilepton final state using proton-proton collisions at √ s = 13 TeV with the ATLAS detector,''
  ATLAS-CONF-2015-070.
  %%CITATION = ATLAS-CONF-2015-070;%%
  %17 citations counted in INSPIRE as of 27 Apr 2016

\bibitem{CMS_dilepton_13TeV}
%\cite{CMS:2015nhc}
%\bibitem{CMS:2015nhc} 
  The CMS Collaboration %[CMS Collaboration],
  %``Search for a Narrow Resonance Produced in 13 TeV pp Collisions Decaying to Electron Pair or Muon Pair Final States,''
  CMS-PAS-EXO-15-005.
  %%CITATION = CMS-PAS-EXO-15-005;%%
  %13 citations counted in INSPIRE as of 27 Apr 2016

%\cite{Huber:2000ie}
\bibitem{Huber:2000ie} 
  S.~J.~Huber and Q.~Shafi,
  %``Fermion masses, mixings and proton decay in a Randall-Sundrum model,''
  Phys.\ Lett.\ B {\bf 498}, 256 (2001).
%  doi:10.1016/S0370-2693(00)01399-X
%  [hep-ph/0010195].
  %%CITATION = doi:10.1016/S0370-2693(00)01399-X;%%
  %379 citations counted in INSPIRE as of 29 Apr 2016

%\cite{Grossman:1999ra}
\bibitem{Grossman:1999ra}
  Y.~Grossman and M.~Neubert,
  %``Neutrino masses and mixings in nonfactorizable geometry,''
  Phys.\ Lett.\ B {\bf 474} (2000) 361.
%  doi:10.1016/S0370-2693(00)00054-X
%  [hep-ph/9912408].
  %%CITATION = doi:10.1016/S0370-2693(00)00054-X;%%
  %753 citations counted in INSPIRE as of 29 Apr 2016

%\cite{Gherghetta:2000qt}
\bibitem{Gherghetta:2000qt} 
  T.~Gherghetta and A.~Pomarol,
  %``Bulk fields and supersymmetry in a slice of AdS,''
  Nucl.\ Phys.\ B {\bf 586}, 141 (2000).
%  doi:10.1016/S0550-3213(00)00392-8
%  [hep-ph/0003129].
  %%CITATION = doi:10.1016/S0550-3213(00)00392-8;%%
  %979 citations counted in INSPIRE as of 29 Apr 2016

%\cite{Georgi:2000ks}
\bibitem{Georgi:2000ks} 
  H.~Georgi, A.~K.~Grant and G.~Hailu,
  %``Brane couplings from bulk loops,''
  Phys.\ Lett.\ B {\bf 506}, 207 (2001).
%  doi:10.1016/S0370-2693(01)00408-7
%  [hep-ph/0012379].
  %%CITATION = doi:10.1016/S0370-2693(01)00408-7;%%
  %226 citations counted in INSPIRE as of 01 May 2016

%\cite{Davoudiasl:2003zt}
\bibitem{Davoudiasl:2003zt} 
  H.~Davoudiasl, J.~L.~Hewett and T.~G.~Rizzo,
  %``Brane localized curvature for warped gravitons,''
  JHEP {\bf 0308}, 034 (2003).
%  doi:10.1088/1126-6708/2003/08/034
%  [hep-ph/0305086].
  %%CITATION = doi:10.1088/1126-6708/2003/08/034;%%
  %35 citations counted in INSPIRE as of 28 Apr 2016

%\cite{Hewett:2016omf}
\bibitem{Hewett:2016omf} 
  J.~L.~Hewett and T.~G.~Rizzo,
  %``750 GeV Diphoton Resonance in Warped Geometries,''
  arXiv:1603.08250 [hep-ph].
  %%CITATION = ARXIV:1603.08250;%%
  %6 citations counted in INSPIRE as of 28 Apr 

%\cite{Hamberg:1990np}
\bibitem{Hamberg:1990np} 
  R.~Hamberg, W.~L.~van Neerven and T.~Matsuura,
  %``A Complete calculation of the order $\alpha-s^{2}$ correction to the Drell-Yan $K$ factor,''
  Nucl.\ Phys.\ B {\bf 359}, 343 (1991); 
  Erratum: [Nucl.\ Phys.\ B {\bf 644}, 403 (2002)].
%  doi:10.1016/0550-3213(91)90064-5
  %%CITATION = doi:10.1016/0550-3213(91)90064-5;%%
  %955 citations counted in INSPIRE as of 13 May 2016

%\cite{Arun:2015kva}
\bibitem{Arun:2015kva} 
  M.~T.~Arun and D.~Choudhury,
  %``Bulk gauge and matter fields in nested warping: I. the formalism,''
  JHEP {\bf 1509}, 202 (2015).
%  doi:10.1007/JHEP09(2015)202
%  [arXiv:1501.06118 [hep-th]].
  %%CITATION = doi:10.1007/JHEP09(2015)202;%%
  %2 citations counted in INSPIRE as of 13 May 2016

\end{thebibliography}
\end{document}